\documentclass[preprint,amsmath,amssymb]{revtex4-1}

\usepackage{graphicx}
\usepackage{dcolumn}
\usepackage{bm}
\usepackage{booktabs}
\usepackage{amsmath}
\usepackage{amssymb}
\usepackage{natbib}
\usepackage{subfigure}
\usepackage{color}

\begin{document}

\title{Van der Waals bonding in layered compounds from advanced first-principles calculations}

\author{T.~Bj\"orkman,$^{1}$,  A.~Gulans,$^{1}$ A.~V.~Krasheninnikov,$^{1,2}$ and R.~M.~Nieminen$^1$}
 \affiliation{$^{1}$COMP - Aalto University School of Science, P.O. Box 11100, 00076 Aalto, Finland}
 \affiliation{$^{2}$Department of Physics, University of Helsinki, P.O. Box 43 00014 Helsinki, Finland}
\date{\today}

\begin{abstract}
Although the precise microscopic knowledge of van der Waals interactions is crucial for understanding bonding in weakly bonded layered compounds, very little quantitative information on the strength of interlayer interaction in these materials is available, either from experiments or simulations. Here, using many-body perturbation and advanced  density-functional theory techniques, we calculate the interlayer binding and exfoliation energies for a large number of layered compounds and show that, independent of the electronic structure of the material, the  energies for most systems are around 20 meV/\AA$^2$.  This universality explains the successful exfoliation of a wide class of layered materials to produce two-dimensional systems, and furthers our understanding the properties of layered compounds in general. 
\end{abstract}
\maketitle

Recent progress in the mechanical\cite{novoselov2005,mos2} and chemical\cite{coleman1,coleman2}  exfoliation of weakly bonded layered inorganic compounds, such as BN, MoS$_2$, WSe$_2$,  Bi$_2$Se$_3$, Bi$_2$Te$_3$, raises  prospects for manufacturing two-dimensional materials which can be used in a plethora of applications\cite{mas-balleste2011}. 
The optimization of the exfoliation process should be helped by a precise knowledge of the interlayer bonding in the parent layered compounds, data which is presently unavailable. 
This lack of data is also hampers the studies of the layered compounds themselves, which can be topological insulators\cite{topins}, thermoelectrics\cite{bi2te3},  charge-density-wave materials\cite{wilson1975} and superconductors \cite{gamble1970}.

Two closely related quantities, the binding energy, $E_B$, between the layers and the energy required to remove an individual layer, the exfoliation energy, $E_{XF}$, are of crucial importance for optimizing the process to produce a two-dimensional structure, as well as for understanding the interlayer bonding in the three-dimensional parent materials. Unfortunately, essentially no information on the interlayer bonding is available from experiments, with the only exception being graphite\cite{girifalco1956,benedict1998,zacharia2004}. Moreover,  the standard first-principles computational approaches based on density-functional theory (DFT) with widely used local and semi-local exchange and correlation (XC) functionals are of little help, since these functionals fail to account for the non-local van der Waals (vdW) interactions between the layers, as has been demonstrated for graphite\cite{harris1985,dobson2001}.

Recently, however, several methodologies that are able to handle vdW interactions have become available for calculations. In this Letter, we apply two of these, the non-local correlation functional method (NLCF) of References \onlinecite{llvdw1,llvdw2,vv10}, and the adiabatic-connection fluctuation-dissipation theorem within the random-phase approximation (RPA)\cite{nozieres_rpa,langreth-perdew1977,rpa1} to study the interlayer binding of layered compounds. The NLCF approach is free from material specific parameters and has been shown to be in good agreement with experimental data for various systems\cite{llvdw1,llvdw2}. 
RPA is expected to be highly accurate in the limit of long wavelength fluctuations involved in the vdW interaction between distant objects\cite{dobson2012} and has served as the basis for analytic vdW theory for a long time\cite{dzyaloshinskii1961}, but is less accurate for short-range interactions involved in the covalent bonding in solids\cite{yan2000,gruneis2009}. 
This has been addressed in a number of recent works aiming to improve the properties of the RPA by introducing further terms in the many-body interaction\cite{gruneis2009,hesselmann2011} and by introduction of approximations to the exchange-correlation kernel\cite{dobson-wang2000}.
However, these extensions come at a formidable computational cost and the short-range deficiency mostly affects the total correlation energy and is less serious when comparing energy differences\cite{yan2000}. The RPA approach has been demonstrated to produce accurate results for small molecules\cite{eshuis2011}, atomization energies in solids\cite{rpa1,rpa2}, surface and adsorption energies\cite{rpa3} and binding of graphite\cite{seb}. In an attempt to get a bird's-eye view of the typical behaviour of the interlayer bonding in weakly bonded layered materials, we perform high-throughput calculations for a large set of compounds, identified by datamining techniques to be likely candidates for layered structures with predominantly vdW type of interactions between the layers. Unfortunately, the RPA is presently prohibitively expensive from the computational point of view to be used as the standard method of choice, and is applied here as a reference for a smaller set of compounds.

A set of layered compounds were selected by searching the Inorganic Crystal Structure Database (ICSD)\cite{icsd} and applying geometric criteria to identify vdW bonded layered structures. 
The criteria were based on the packing ratio of the crystal, identification of gaps in the structure along the crystallographic $c$ axis, and verification that the interlayer bonds were elongated beyond what is expected for covalent bonds by comparison with the sum of the covalent radii. The filtering procedure is described in detail in the Supplementary Material.
The benefit of this procedure is that it will generate a selection unbiased by our own expectations and previous knowledge, thus providing a more diverse set. From the generated list of compounds, we selected a subset of tetragonal and hexagonal/trigonal systems and further enlarged the list by making sure that all reported layered compounds of transition metal dichalcogenides (MX$_2$, with M being a transition metal and X being either S, Se or Te) were in the list. After removal of some of the most computationally demanding systems, we arrive at a mixed set of 86 compounds -- metals, semimetals, insulators, magnetic compounds. Apart from the MX$_2$ compounds, the list thus obtained contains many important materials, such as graphite, BN and the topological insulators Bi$_2$Se$_3$ and Bi$_2$Te$_3$. 
All calculations were performed using the projector augmented wave method as implemented in the electronic structure package VASP\cite{vasp3,kresse1999}. We used an in-house NLCF implementation\cite{andrisvdw} and the standard VASP implementation of RPA\cite{rpa1}. Crystal geometries were automatically generated from database searches using the program CIF2Cell\cite{cif2cell}.

\begin{figure}[htbp] 
   \centering
   \includegraphics[width=0.95\textwidth]{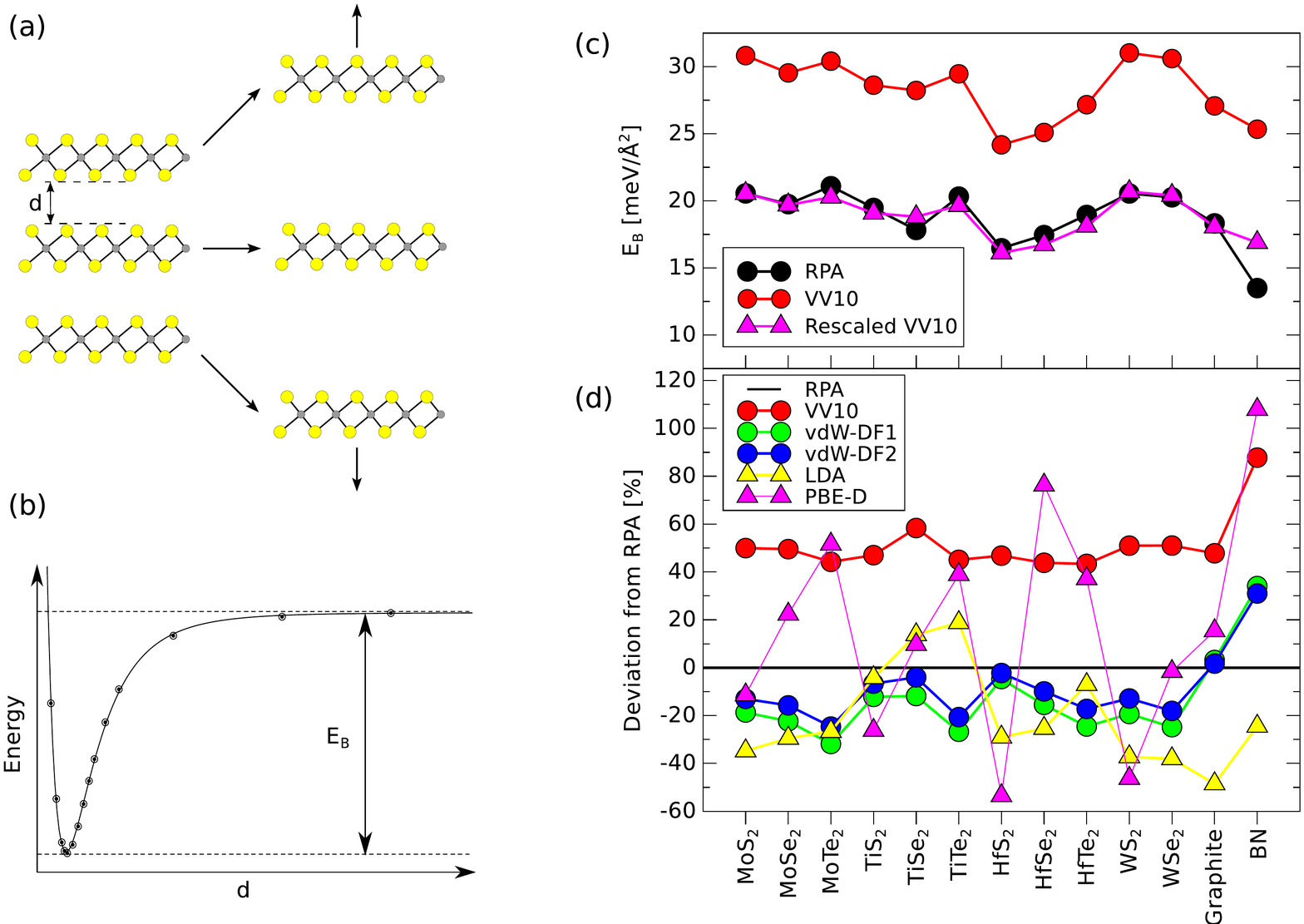}
   \caption{(a)  Procedure of calculating the interlayer binding energy by increasing interlayer distance, $d$. (b) Schematic illustration of a binding energy curve. (c) A set of interlayer binding energies calculated using the RPA and the VV10 functional, demonstrating how a rescaling of the VV10 values can be used to match the more computationally demanding RPA values. (d) Comparison chart for a number of different functionals widely used for treating vdW interactions relative to the RPA results.}
   \label{fig1}
\end{figure}

The procedure for calculating $E_B$ is illustrated schematically in Figure \ref{fig1}(a,b). In order to get accurate estimates of $E_B$, a general assessment of the different DFT-based approaches was necessary. The list of investigated methods included the local density approximation (LDA), the semi-empirical method by Grimme\cite{Grimme} (PBE-D) as well as the NLCF methods by Dion et al.\cite{llvdw1} (vdW-DF1), Lee et al.\cite{llvdw2} (vdW-DF2) and Vydrov and van Voorhis\cite{vv10} (VV10). We compared the calculated interlayer binding energies to the more sophisticated many-body treatment of RPA for a subset of layered compounds, and studied how well the different DFT-based approaches reproduce the reported vdW bond lengths, the only experimental data pertaining to the vdW interaction that is available for all compounds. 

The conclusion is that all NLCF methods reproduce the RPA trends of $E_B$ sufficiently well to be useful for predicting interlayer binding energies, whereas two other popular choices for treating vdW interactions, LDA and PBE-D, do not\footnote{Note however, that despite their less good performance for binding energies, both LDA and PBE-D typically produce equilibrium geometries in good agreement with experimental data.}. In fact, any of the NLCF type of functionals can be rescaled by its average deviation from the calculated RPA values to yield an estimate of the RPA energy, limited primarily by the inaccuracies in equilibrium bond lengths.
In particular, we find that the VV10\cite{vv10} functional is highly successful, both for producing accurate geometries and following the $E_B$ trends of RPA very closely, so that an accurate estimate of the RPA binding energy can be obtained by simply rescaling the VV10 results by a factor of 0.66, and we will henceforth refer to this as the \emph{NLCF estimate} of the binding energy. It should be noted that this estimate is purely based on the empirical observation of the trends for the 28 compounds investigated by RPA.
This is illustrated in Figure \ref{fig1}, where a representative set of the calculations are shown, first as a demonstration of the effect of the rescaling in panel (c), and then by a comparison of a set of different functionals in terms of their relative deviations from RPA in panel (d). We also point out that RPA is superior to all other methods at reproducing the experimental vdW interplanar bond lengths, with a maximal deviation of 4\%, thus further demonstrating the high accuracy of RPA for vdW bonding in layered compounds. The full data set used for the functional comparison is available in the Supplementary material, Sections II and IV.

\begin{figure}[htbp] 
   \centering
   \includegraphics[width=0.75\textwidth]{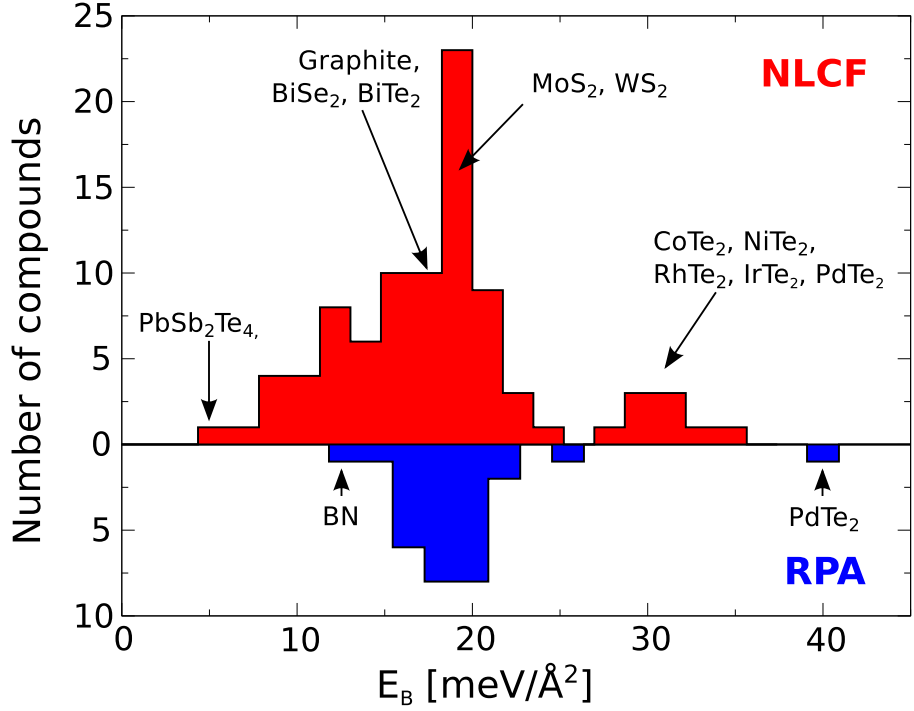}
   \caption{Distribution of binding energies estimated from a NLCF (VV10), and distribution of the binding energies calculated by RPA in blue. The vast majority of the compounds fall in the interval $\sim13-21$ meV/\AA$^2$. We also mark in which histogram bin some particular compounds are. The outliers on the high binding energy side around 30 meV/\AA$^2$ are mostly a set of tellurides where weak covalent bonds contribute as well.}
   \label{bindendist}
\end{figure}

The smaller set of $E_B$ calculated using RPA and the full set estimated by rescaling of the VV10 data, are shown in Figure \ref{bindendist}. The peak of the distribution is around 13-21 meV/\AA$^2$ (taken as one standard deviation around the average of the distribution), with a slightly more significant tail towards lower than towards higher binding energies. This region contains, among other compounds, graphite and BN, and also most of the MX$_2$ compounds. There are outliers in the distribution at slightly higher binding energies, consisting primarily of the Co family ditellurides and NiTe$_2$ and PdTe$_2$. These compounds have significant binding energies (15-25 meV/\AA$^2$) even when calculated using a regular generalized-gradient approximation (GGA) functional, which normally produces little or no binding for vdW-bonded systems. This indicates that, although there are  contributions also from covalent interactions captured by the GGA type functional, in a few cases, the size of the vdW component of the binding remains the same. 
We have not been able to find correlations of $E_B$ to any other quantity in the present set of compounds. The quantities scanned for such correlations were the interlayer distances, intralayer thicknesses and band-gap/metallicity as well as properties of the constituent atoms such as the atomic weights and polarizabilities. Nor can we find any reason such as simple band filling arguments that would give any correlation to the binding energies.
We conclude that the strength of the vdW bonds in layered solids is a universal quantity. Such a universality is in line with observations by Coleman et al.\cite{coleman1,coleman2}, based on the experimental data on chemical exfoliation of a large set of MX$_2$ and Bi$_2$Te$_3$ compounds. Detailed information on the binding energies for specific compounds is tabulated in the Supplementary Material, Section IV.  

The statement of universality of the vdW component of the binding energy of layered compounds raises the question whether our initial selection criteria might have been biased in such a way that we only find compounds with a vdW component of the binding energy in this range. Within a given selection it is of course never possible to validate the selection itself, but we nevertheless gain confidence by the lack of correlation to any conspicuous quantity within our selection. It is hard to see how one could arrive at some group of compounds with different binding properties in such a way that it does not constitute a variation of some the properties to which we have found no correlation within our data set.

In view of the known qualitative differences between metals and semiconductors for large separations\cite{dobson2006}, the observed universality seems counterintuitive, but can be understood through simple arguments. The binding energy is determined by the balance of the repulsive and attractive parts of the interaction near the equilibrium geometry, and these quantities depend on the electron density profile. The repulsive part stems from the exchange interactions and can be estimated well based on the electronic density alone\cite{murray2009}. Similar considerations apply to the attractive vdW interactions, described e.g. by Zaremba and Kohn\cite{zarembakohn}, who derived a form for the high frequency -- long wavelength limit of the density response of  a surface in terms of the density profile, and were also among the arguments leading up to the original formulation of the NLCF method\cite{rydberg2003}. As the density profiles of different vacuum interfaces show similar exponential decays, we can understand why the vdW component of the binding is constant and larger variations come from covalent bonding.

\begin{figure}[htbp]
   \centering
   \includegraphics[width=0.75\textwidth]{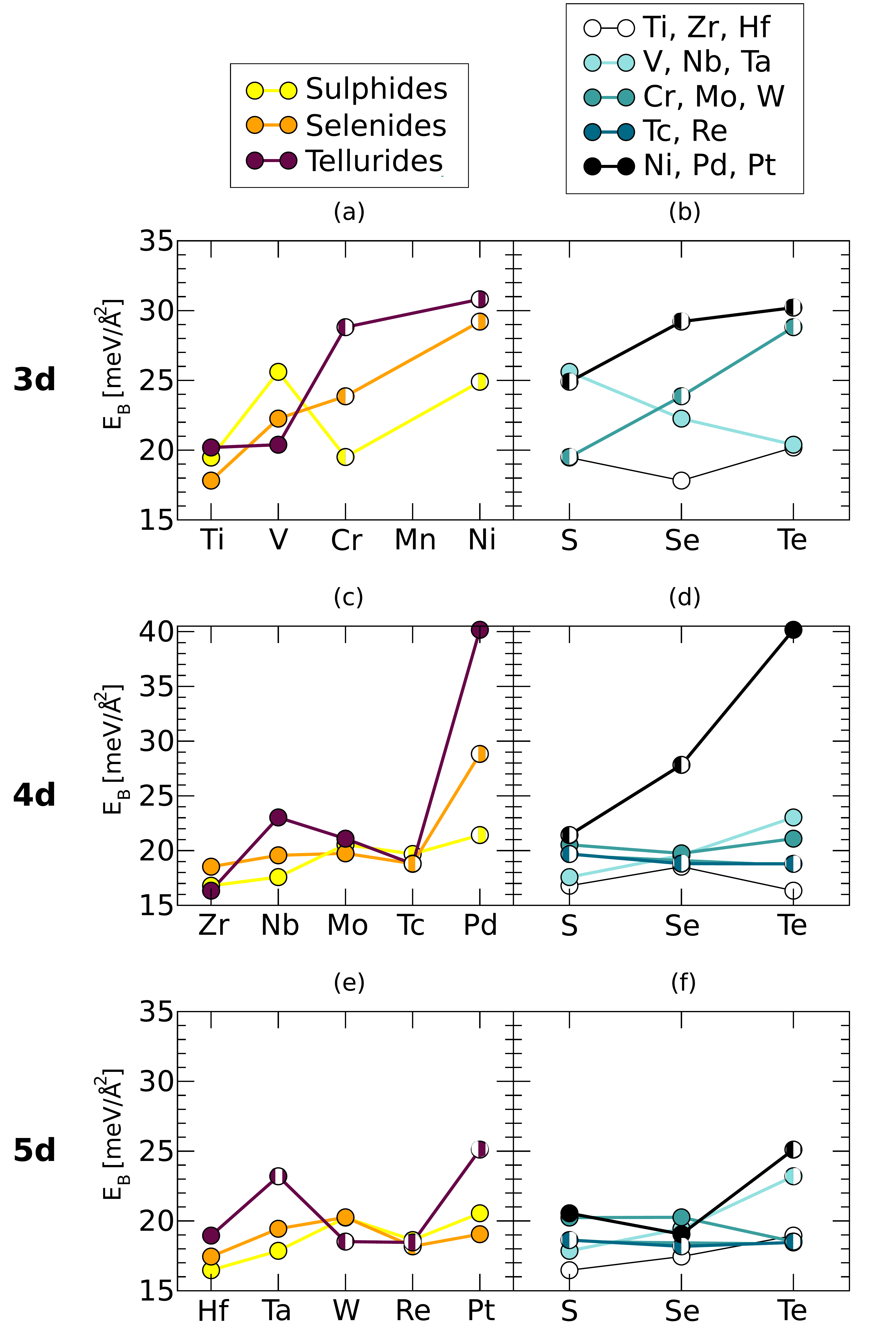}
   \caption{Interlayer binding energies of the transition metal dichalcogenides. Panels (a,c,e) show the variation of the binding energy with respect to the transition metal species and panels (b,d,f) show the variation of the binding energy with respect to the chalcogen species. The rows stand for the $3d$, $4d$ and $5d$ transition metal series, respectively. Solid circles are the results of the RPA calculations and a striped pattern indicate a value obtained from the  NLCF (VV10) calculations.}
   \label{dichalcogenide_binding_energies}
\end{figure}

Taking into account the recent interest in layered MX$_2$ systems\cite{coleman1,coleman2} we present in Fig. \ref{dichalcogenide_binding_energies} $E_B$ for all  layered forms of MX$_2$ compounds, which are found in the early and late transition metal $d$ series. We have also filled out some gaps among the experimentally reported structures by calculations for hypothetical layered structures of CrTe$_2$, TcSe$_2$, TcTe$_2$, ReTe$_2$, NiS$_2$ and NiSe$_2$. The crystallographic parameters for these compounds are reported in the Supplementary Material, Section III. Our findings are shown in Figure \ref{dichalcogenide_binding_energies}, illustrating the variation of $E_B$ as we move across the transition metal series, and by the respective chalcogen species. Most energies fall in the region $E_B=$15-20 meV/\AA$^2$ and, as a rule, the factor that most strongly determines the binding energy appears to be the transition metal species, while the dependency on the chalcogen species is weaker. Exceptions to these rules are found among the Cr compounds and the Co and Ni family tellurides, which, as previously discussed, have large covalent and electrostatic contributions to the binding energies. Inasmuch as the atomic polarizabilities vary smoothly as function of the transition metal or chalcogen species\cite{nagle1990}, the lack of persistent trends in Figure \ref{dichalcogenide_binding_energies} is important. This demonstrates the importance of a correct description of the electronic states, incorporating collective effects such as the band formation, to capture trends in the binding energies.

\begin{figure}[htbp] 
   \centering
   \includegraphics[width=0.75\textwidth]{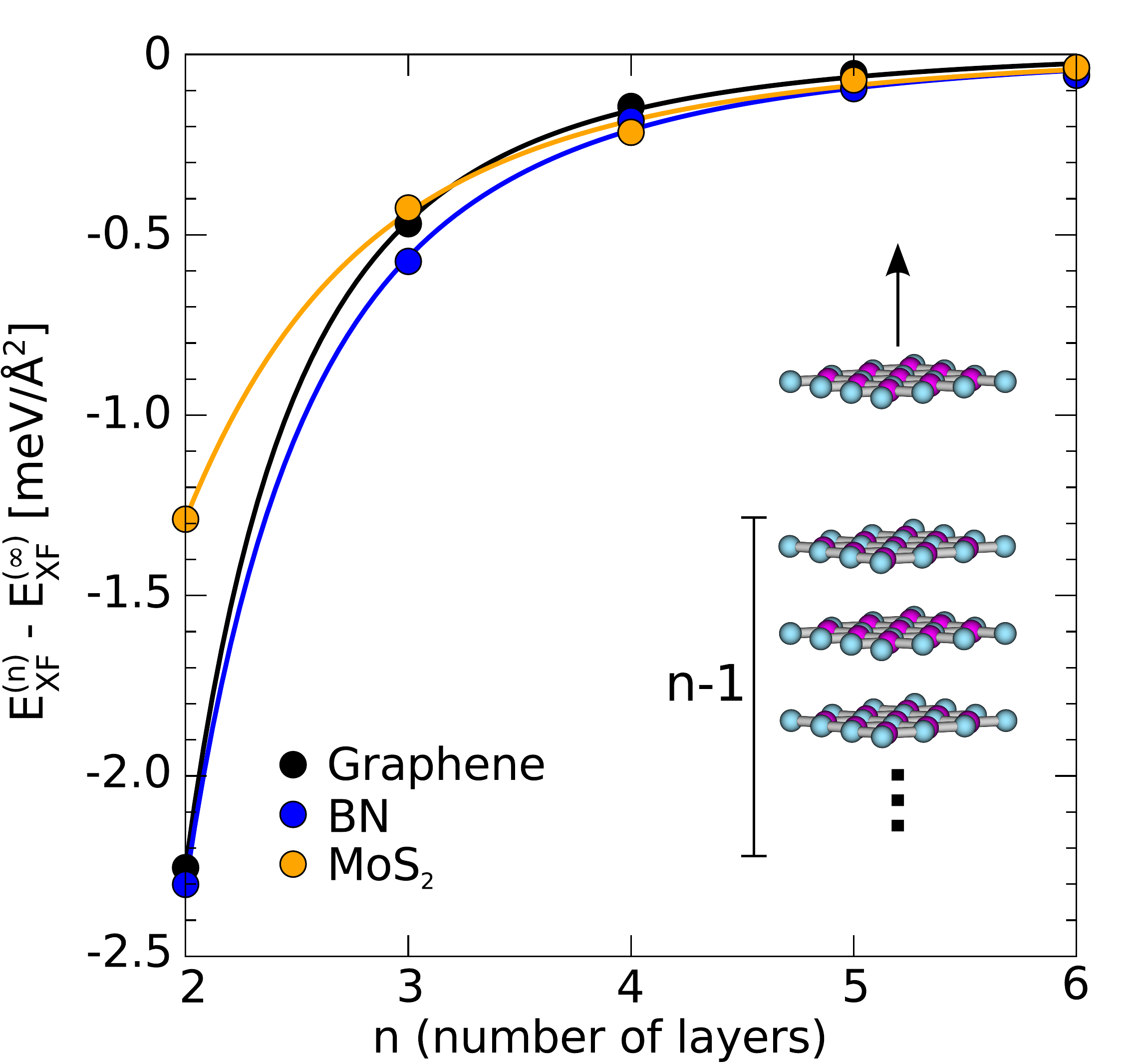}
   \caption{Energy required for exfoliation of a single layer from a multilayer structure as function of the number of layers, $n$, as shown schematically in the inset figure. Curves has been fitted to the calculated points as a guide for the eye and the zero of energy has been set to $E^{(\infty)}_{XF}$,  the asymptote of the respective curve as $n$ is taken to infinity. The energy is estimated from NLCF (VV10) calculations.}
   \label{multilayers}
\end{figure}

The interlayer binding energy is closely related to the exfoliation energy, $E_{XF}$, the cost of removing a single layer from the surface of the bulk compound. It is expected that $E_{XF}\approx E_B \approx 2\cdot E_{surf}$, where $E_{surf}$ is the surface energy, and this point is further explained in Section IV of the Supplementary Material. We simulated exfoliation for a series of multi-layer systems by peeling off the top layer, as shown in the inset of  Figure \ref{multilayers}. The figure demonstrates, for the cases of graphene, BN and MoS$_2$, how peeling off a single layer costs increasingly large amounts of energy as the number of underlying layers increase. This behaviour originates from  the interaction of the topmost layer with not only its nearest neighbour, but also other layers. However, the difference between  $E_B$  and $E_{XF}$ is small, no more than 4\%, primarily due to surface relaxation effects, as our calculations for graphene, BN,  and all hexagonal, non-magnetic MX$_2$ compounds indicate. Thus, the exfoliation energy can be assumed to be equal to the interlayer binding energy in all layered materials, so that our accurate theoretical results for interlayer binding energies are not only important for understanding the properties of bulk layered compounds and inorganic multi-walled nanotubes\cite{nagapriya2008}, but should also be useful in the optimization of the exfoliation process.

In conclusion, using advanced calculation techniques we have shown that the interlayer binding energies of weakly bonded layered compounds are found in a small energy interval of 13-21 meV/\AA$^2$. These energies fall very close to the exfoliation energies of the compounds, and are of high importance for the understanding of weakly bonded layered solids and their exfoliation into single layers.

This research was supported by the Academy of Finland through the COMP Centre of Excellence Grant 2006-2011. Computational resources were provided by Finland's IT center for Science (CSC). The ISCD has kindly granted permission to export and store the database in a format appropriate to our purpose.

%

\end{document}


\title{Supplementary material}
\author{T. Bj\"orkman}
 \affiliation{COMP - Aalto University School of Science, P.O. Box 11100, 00076 Aalto, Finland}
\author{A.~Gulans}
 \affiliation{COMP - Aalto University School of Science, P.O. Box 11100, 00076 Aalto, Finland}
\author{A.~V.~Krasheninnikov}
 \affiliation{COMP - Aalto University School of Science, P.O. Box 11100, 00076 Aalto, Finland}
 \affiliation{Department of Physics, University of Helsinki, P.O. Box 43 00014 Helsinki, Finland}
\author{R.~M.~Nieminen}
 \affiliation{COMP - Aalto University School of Science, P.O. Box 11100, 00076 Aalto, Finland}
\date{\today}

\maketitle

\section{Methods}

\subsection{Compound selection}
We wish to identify layered compounds with interlayer bonding dominated by van der Waals (vdW) interactions and for computational reasons we wish these to have as high symmetry as possible.
We selected the compounds in the study by successive filtering of all the compounds in the Inorganic Crystal Structure Database (ICSD). The strategy is to identify large deviations from the expected behaviour of covalently, ionically or metallically bonded solids, and a primary tool is the set of covalent radii of the compounds. We choose to simplify the problem by restricting the search to layered compounds where the layers are perpendicular to the crystallographic $c$ axis, since the compounds we are interested in have a unique crystallographic axis and high in-plane symmetry perpendicular to this axis, and such compounds are conventionally chosen to have the $c$ axis as the unique axis. Compounds that do not conform to this symmetry criterion are discarded. In a second coarse step, we filter out compounds based on the packing ratio, defined as the covalent volume divided by the total cell volume. This filters out systems such as close-packed metallic systems and very open molecular solids. In the third step we identify large gaps in along the crystallographic $c$ axis, indicating that there may be layers bonded primarily by vdW forces. In the fourth and last step we select only those structures with a gap such that the distance between neighbouring atoms across the gap is significantly larger than the sum of their covalent radii.
Several different sets of covalent radii were tested and the precise choice was found not to be important for the identification of the very overstretched bonds of primarily vdW bonded solids.

For computational reasons we also need to remove all non-stoichiometric compounds and compounds with overly large unit cells. We also remove any compound containing $f$-electron elements, since for the present purposes we wish to avoid all additional complications arising from the strong-correlation physics involved in these compounds. The study of binding energies also includes any of the layered transition metal dichalcogenides, also those previously discarded by the symmetry criteria.

\subsection{Electronic structure calculations}

We used the projector-augmented wave (PAW) potentials  from the library distributed with the VASP code\cite{kresse1999}.
The plane wave cutoff was initially selected as 1.5 times the default cutoff, which was subsequently increased in individual cases if there were apparent convergence problems. The convergence was also more carefully tested for a small subset of compounds. Compounds containing elements in the $3d$ series from Cr to Ni were calculated in the ferromagnetic mode. 
Brillouin zone integrations were performed using the Gaussian smearing method with a smearing width of 0.1eV, using a uniform Monkhorst-Pack $k$-point mesh with the number of points selected to give a distance of 0.2\AA$^{-1}$ between the mesh points for non-magnetic calculations and 0.15\AA$^{-1}$ for magnetic calculations. 

The RPA correlation energy was calculated using the adiabatic connection-fluctuation dissipation theorem. We used the standard VASP implementation\cite{harl2008}, where the density-response function is represented in the plane-wave basis. The size of the basis is characterized by the energy cut-off, which has a strong influence on the correlation energy. In particular, Harl and Kresse have suggested that the correlation energy converges as
\begin{equation}
\label{convergence1}
E_\mathrm{c}^\mathrm{RPA}(q)=E_\mathrm{c}^\mathrm{RPA}(q=\infty)+A/q^3,
\end{equation}
where A is a constant and $q$ is the cut-off wavenumber that can be related to the cut-off energy through the relation $E_\mathrm{cut-off}=q^2/2$. However, it can be shown\cite{rpa1} that Eq.~\ref{convergence1} can be extended to 
\begin{equation}
\label{convergence2}
E_\mathrm{c}^\mathrm{RPA}(q)=E_\mathrm{c}^\mathrm{RPA}(q=\infty)+A/q^3+B/q^5+C/q^6+\cdots,
\end{equation}
where A, B and C are constants. However, we find numerically that when the energy differences involved in the vdW binding energies are calculated, the terms containing $q^{-3}$ and $q^{-6}$ vanish. Then, for energy differences the following relation holds
\begin{equation}
\label{convergence3}
\delta E_\mathrm{c}^\mathrm{RPA}(q)\approx \delta E_\mathrm{c}^\mathrm{RPA}(q=\infty)+\alpha/q^5+\beta/q^7+\cdots,
\end{equation}
where $\alpha$ and $\beta$ are constants. In practical calculations, we have calculated RPA correlation energies using different cut-off energies and have used them for fitting of Equation~\ref{convergence3}. This procedure allowed us to obtain accurate estimates of the complete basis limit with cut-off energies as low as 100--150~eV, which are significantly lower than those previously used in Refs. \onlinecite{harl2008,rpa1}. Translated into computational effort, this procedure allows us to obtain the binding energies cheaper by an order of magnitude without sacrificing the accuracy.

The total RPA energy was evaluated as a sum of the correlation energy and the total energy from a non-self-consistent exact exchange calculation. In both cases, PBE orbitals were used.

All systems were studied at their experimental in-plane lattice constant, but at the equilibrium interlayer spacing for the different methods, with the exception of the hypothetical compounds presented in Section \ref{hypotheticalcompounds}, where the in-plane lattice constant was relaxed using the VV10 functional. The input structure to the electronic structure program was generated by stretching of the $c$-axis with the layers intact. The atoms were then allowed to relax to their equilibrium positions with a fixed unit cell so that the intralayer geometry was automatically relaxed, and the interlayer geometry was relaxed by hand, so as to obtain a binding energy curve as a function of the $c$-axis length as shown in Figure 1 (b) of the main paper. By fitting the total energy points closest to the minimum to a polynomial, we obtained values for the equilibrium length and the $C_{33}$ elastic constant. RPA calculations were done with fixed layers, only varying the interlayer distance. The effect of this approximation on the binding energy was found to be negligible, but there is a softening of the $C_{33}$ elastic constant of about 10\% when the layers are relaxed. 
Supercells for calculation of exfoliation energies were constructed by stacking 6 layers of the compound and then adding 6 layers of vacuum. Layers were then removed one by one and the energy difference between 6 and 5+1 layers were calculated. Supercell convergence was tested with respect to the number of layers and the size of the vacuum region.

\section{Investigation of non-local correlation functionals}

In addition to the functional listed in the main paper, we investigated the Perdew-Burke-Enzerhof (PBE)\cite{pbe} GGA functional and the effect of applying the NLCF of Dion et al. on top of the PBE exchange functional (vdW-DF1 (PBE)). The functionals were tested in two different ways, by comparison of relaxed geometry to experiment and by comparison of interlayer binding energies to RPA. 

\begin{figure}[htbp] 
   \centering
   \includegraphics[width=0.95\textwidth]{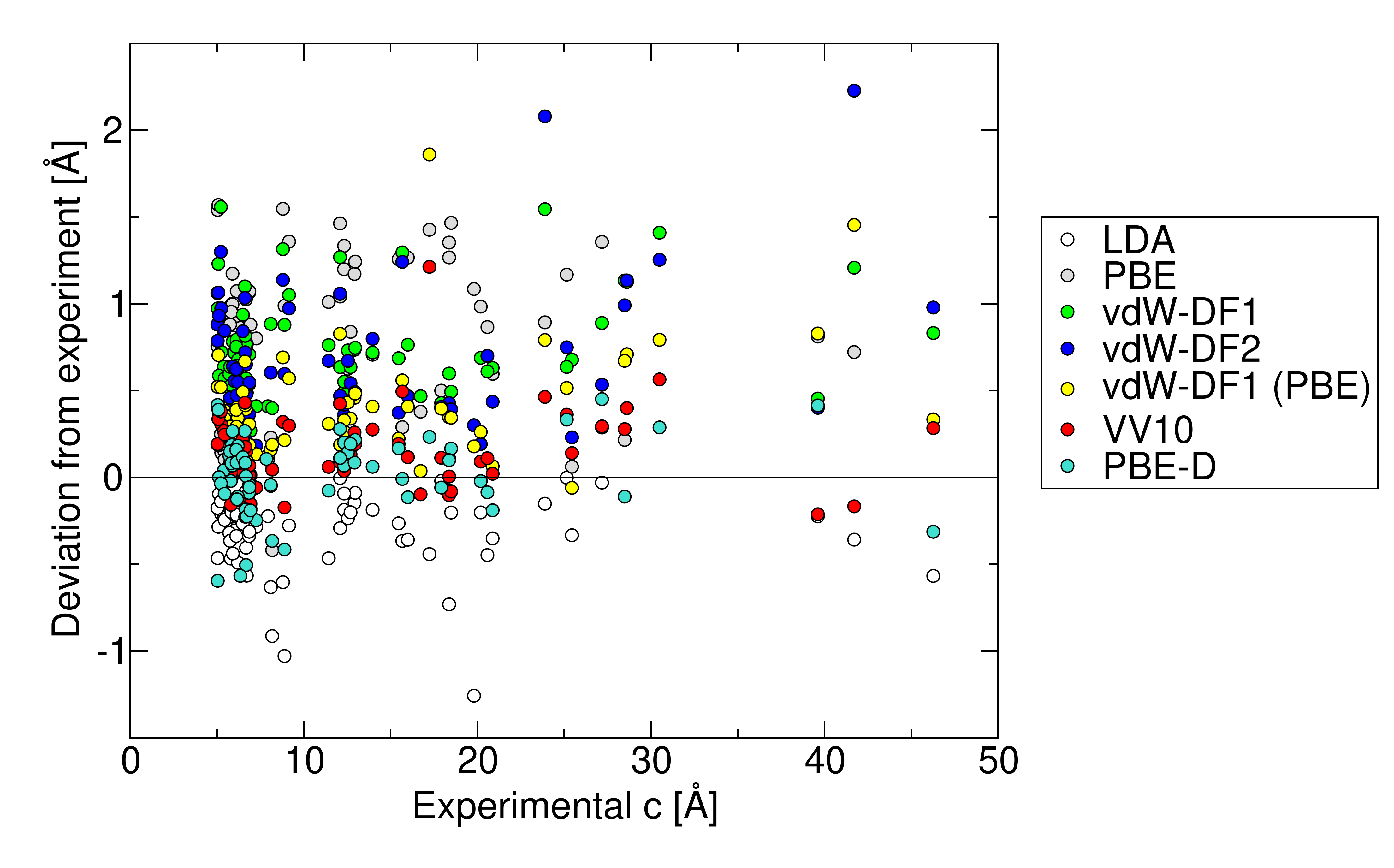}
   \caption{Deviations (\AA) of the calculated $c$-axis values from the experimentally reported value.}
   \label{expdev}
\end{figure}

\subsection{Geometrical considerations}
For the structural benchmarks, structures of 72 compounds were calculated using various NLCF's, and 28 compounds were calculated with RPA. Figure \ref{expdev} shows the deviations from the experimental $c$-axis length for all NLCF and standard LDA/GGA functionals tested. Table \ref{expdevtab} summarizes the deviations from the experimental $c$-axis lengths in terms of: mean relative deviation (MRD), mean absolute relative deviation (MARD), maximal absolute relative deviation (Max. ARD) and one standard deviation (Sigma RD). It can immediately be seen that RPA performs significantly better than any other method for the smaller set of compounds where it was applied. The VV10 functional performs significantly better than other density functionals. The PBE-D functional reproduces well geometrical properties, but it shows large maximal deviations, illustrating a tendency of sometimes failing badly for no obvious reason, and for 6 systems it completely failed to give a sensible binding energy curve. The failures are mostly for systems that contain heavier elements, which are likely to have the least well-tested parameters. It seems probable that suitable refitting of the empirical parameters of PBE-D for layered systems would yield much better results. LDA also performs rather well for structural properties, giving numbers that are in line with the normal LDA overbinding for covalent bonds. The two functionals vdW-DF1 and vdW-DF2 are barely improvements over the PBE result for the $c$-axis lengths, but the shape of the binding energy curves are very different, and so are the $C_{33}$ elastic constants. Applying the original van der Waals density functional on top of PBE instead of RPBE gives a significant improvement of the structural properties. The reason for this is that the spurious LDA binding remains to a larger extent in the PBE functional than in RPBE and this improves the van der Waals bond lengths.

\begin{table}[htdp]
\caption{Deviations from experimental values of the different functionals investigated.}
\begin{center}
\begin{tabular}{l|cccc}
Method & MRD [\%] & MARD [\%] & Max dev. [\%] & $\sigma$ [\%] \\\hline
RPA & 1.0 & 1.5 & 4.5 & 1.6 \\
VV10 & 1.5 & 2.0 & 7.3 & 2.2 \\
vdW-DF1& 8.5 & 8.5 & 29.8 & 5.5 \\
vdW-DF2 & 7.2 & 7.2 & 24.9 & 5.5 \\
vdW-DF1 (PBE) & 3.9 & 3.9 & 13.9 & 3.0 \\
PBE-D & 0.1 & 2.1 & 11.8 & 3.1 \\
LDA & -3.5 & 3.5 & 11.6 & 2.6 \\
GGA & 9.5 & 9.6 & 30.9 & 6.5
\end{tabular}
\end{center}
\label{expdevtab}
\end{table}%

\begin{figure}[htbp] 
   \centering
   \includegraphics[width=0.95\textwidth]{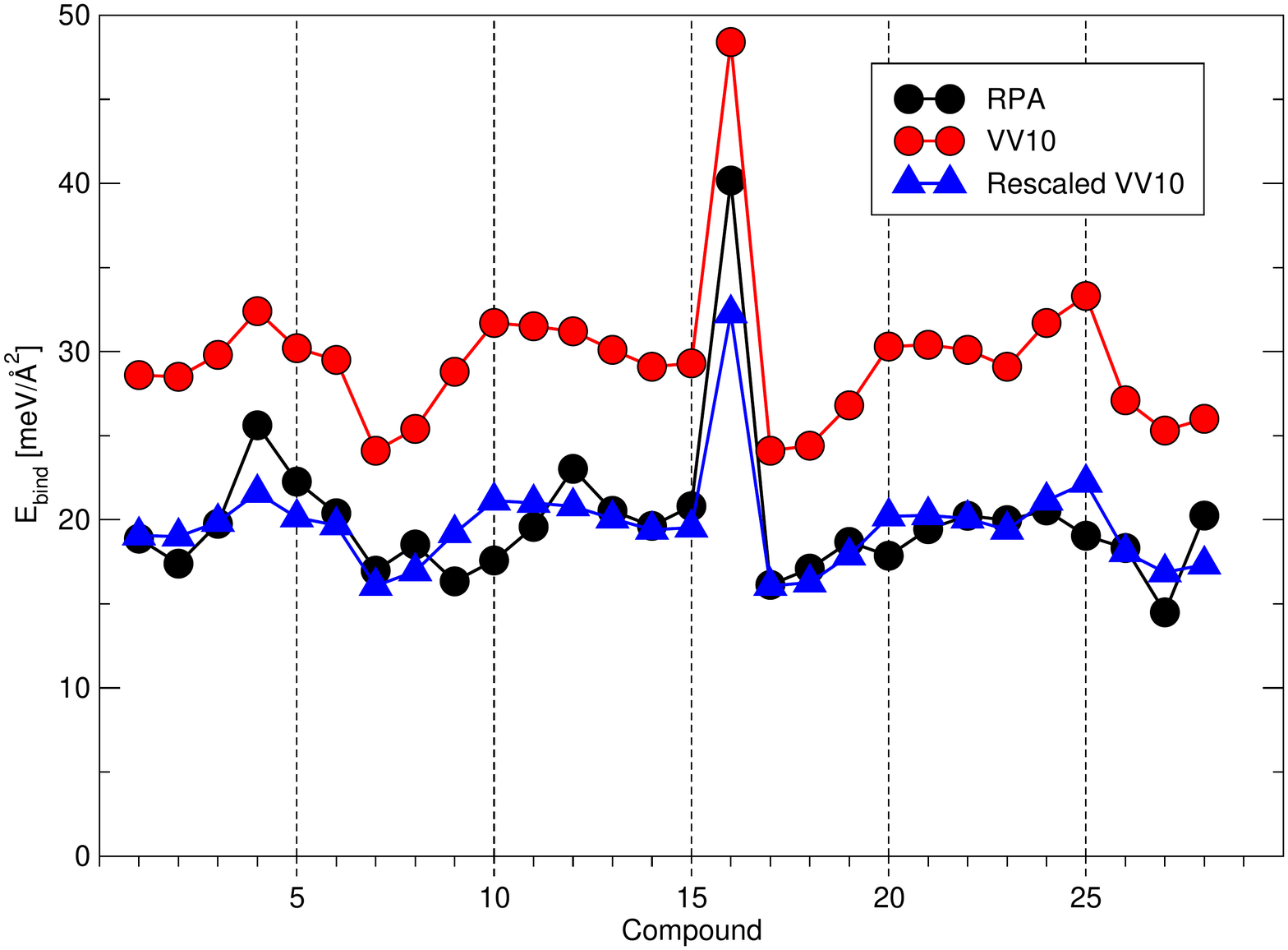}
   \caption{Interlayer binding energies for a set of compounds calculated with RPA and VV10. Also shown is the result of rescaling the VV10 binding energies by a factor of 0.66. The numbering of the compounds can be found in Table \ref{rpaenergies}.}
   \label{rpacomp}
\end{figure}

\subsection{Energy considerations}
To get a high-quality energy benchmark, RPA calculations of binding energies were performed for 28 materials, in addition to the previously published value of graphite by Leb{\`e}gue et al.\cite{seb}.  The results of the comparison with the other functionals, shown in Figure 1 of the main paper, demonstrates that the vdW-DF type of functionals follows the trends of the RPA calculations well. The VV10 functional follows the trends of the RPA calculations particularly well. If rescaled by a factor of about 2/3, the VV10 energies very closely follow the RPA energies, as shown in Figure \ref{rpacomp} for all compounds investigated by RPA in the present study.  In view of Figure 1 of the main paper, one could argue that the Langreth-Lundqvist functionals are in fact closer to the RPA results. However, the failure of these functionals to produce the correct van der Waals bond lengths produces anomalies for the outliers of the interlayer binding energy distribution, the Co family ditellurides and NiTe$_2$ and PdTe$_2$, as discussed in the main paper. For this reason, we prefer to use the VV10 functional to estimate the interlayer binding energies.

\clearpage
\newpage

\section{Crystal structures of hypothetical compounds}\label{hypotheticalcompounds}

We performed relaxations of the crystal structures of layered MX$_2$ compounds not found in the literature. To more completely investigate the trends in binding energies, we also included NiS$_2$ and NiSe$_2$, which experimentally are found in the pyrite structure, but which where found to be electronically stable in the layered $P\bar{3}m1$ structure, which indicates that they might also be possible to synthesize in layered structures. The relaxed cell parameters are given below in Table \ref{crystparms}.

\begin{center}
\begin{longtable}{ll|c|c|c|c|c|c}
\caption[Crystallographic parameters of hypothetical layered compounds]{Crystallographic parameters of hypothetical layered compounds calculated using the VV10 functional. M refers to the transition metal and X to the chalcogenide and units for $a$, $b$ and $c$ are \AA, angles are in degrees and the wyckoff positions are given.} \label{crystparms}
\\
& &  TcSe$_2$ ($P\bar{1}$) & TcTe$_2$ ($P\bar{1}$ ) & ReTe$_2$ ($P\bar{1}$ ) & CrTe$_2$ ($P\bar{3}m1$) & NiS$_2$ ($P\bar{3}m1$) & NiSe$_2$ ($P\bar{3}m1$) \\\hline\hline
\endfirsthead 

\multicolumn{8}{c}
{{\bfseries \tablename\ \thetable{} -- continued from previous page}} \\
& &  TcSe$_2$ ($P\bar{1}$) & TcTe$_2$ ($P\bar{1}$ ) & ReTe$_2$ ($P\bar{1}$ ) & CrTe$_2$ ($P\bar{3}m1$) & NiS$_2$ ($P\bar{3}m1$) & NiSe$_2$ ($P\bar{3}m1$) \\\hline\hline
\endhead

& $a$ & 6.77 & 7.17 & 7.18 & 3.59 & 3.42 & 3.65 \\
& $b$ & 6.89 & 7.33 & 7.32 & 3.59 & 3.42 & 3.65 \\
& $c$ & 6.97 & 7.18 & 7.27 & 6.74 & 4.68 & 5.00 \\
& $\alpha$ & 63.85 & 92.10 & 92.01& 90 & 90 & 90 \\
& $\beta$ & 103.87 & 105.05 & 104.95 & 90 & 90 & 90 \\
& $\gamma$ & 118.91& 118.96 & 118.99 & 120 & 120 & 120\\
M1 & $x$ & 0.067 & 0.069 & 0.072 & 0 & 0 & 0 \\
&  $y$ & 0.313 & 0.321 & 0.322 & 0 & 0 & 0 \\
& $z$ & 0.511 & 0.492 & 0.491 & 0 & 0 & 0 \\
M2 & $x$ & 0.488 & 0.487 & 0.486 & -- & -- & -- \\
& $y$ & 0.290 & 0.297 & 0.298 & -- & -- & -- \\
& $z$ & 0.504 & 0.496 & 0.495 & -- & -- & -- \\
X1 & $x$ & 0.365 & 0.245 & 0.245 & 1/3 & 1/3 & 1/3 \\
& $y$ & 0.307 & 0.380 & 0.382 &  2/3 & 2/3 & 2/3 \\
& $z$ & 0.803 & 0.197 & 0.195 & 0.250 & 0.249 & 0.248\\
X2 & $x$ & 0.824 & 0.717 & 0.717 & -- & -- & -- \\
& $y$ & 0.276 & 0.333 & 0.333 & -- & -- & -- \\
& $z$ & 0.772 & 0.233 & 0.235 & -- & -- & -- \\
X3 & $x$ & 0.140 & 0.225 & 0.224 & -- & -- & -- \\
& $y$ & 0.170 & 0.107 & 0.107 & -- & -- & -- \\
& $z$ & 0.269 & 0.723 & 0.723 & -- & -- & -- \\
X4 & $x$ & 0.679 & 0.753 & 0.755 & -- & -- & -- \\
& $y$ & 0.188 & 0.136 & 0.138 & -- & -- & -- \\
& $z$ & 0.296 & 0.698 & 0.695 & -- & -- & -- 
\end{longtable}
\end{center}

\section{Binding and exfoliation energies}

Here a brief explanation of the different energies discussed in the text, we tabulate the dichalcogenide binding energies shown in Figure 3 of the main paper, as well as all energies calculated using RPA. Last, we list the results of all the different functionals investigated, giving $E_B$, the $C_{33}$ lattice constant and the $c$ axis lengths and their deviation from experimental values. The $c$ axis deviations are given as an interval calculated from the range of different values found in the ICSD database.

\subsection{Relations between different energies discussed in the paper}

\begin{figure}[htbp] 
   \centering
   \includegraphics[width=0.84\textwidth]{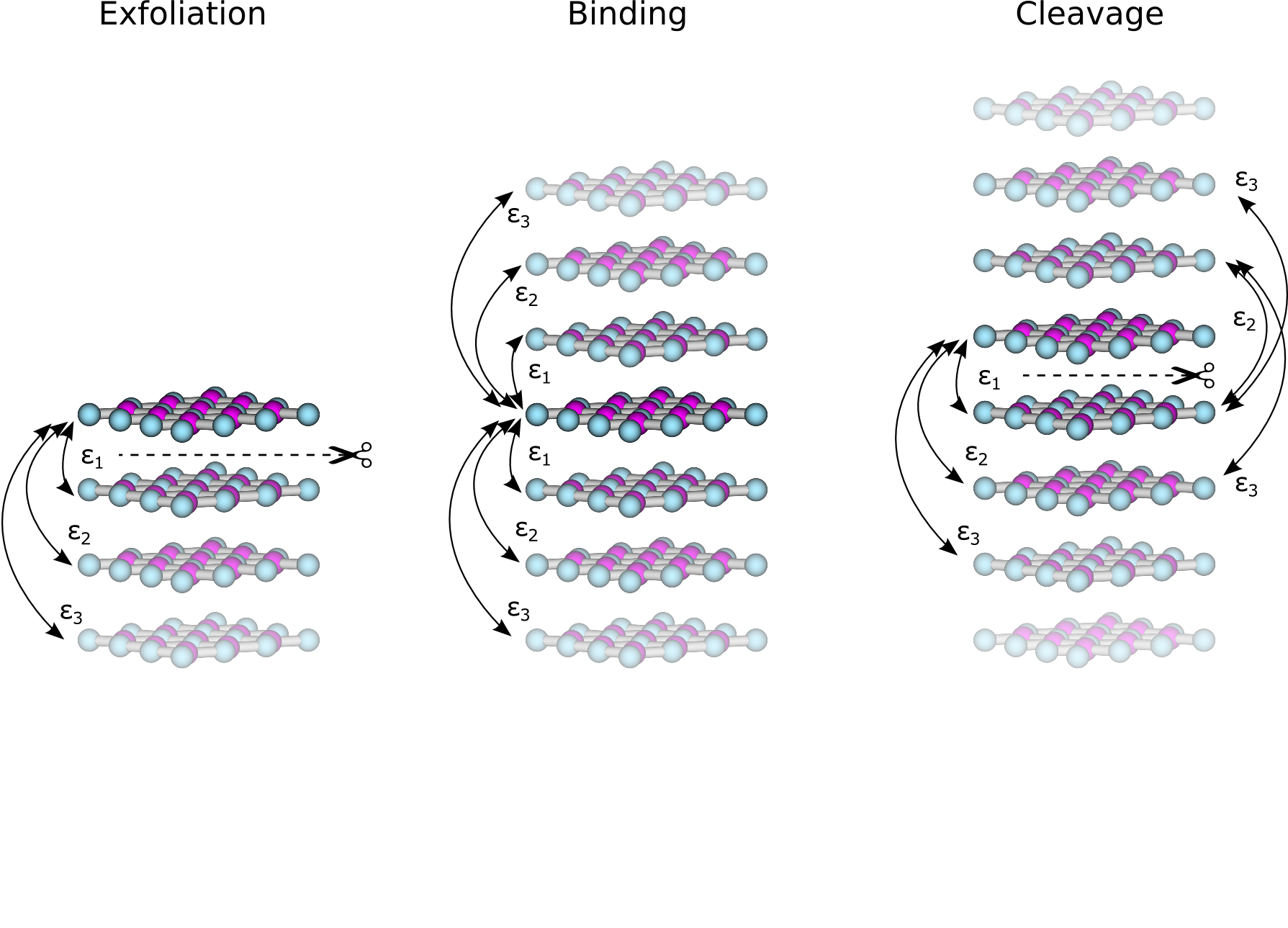}
   \caption{The bonds involved in the binding of layers in the solid. The layer under consideration is solid, and layers get more transparent as we move away from this layer, to illustrate that each layer only are affected by its nearest neighbours. The dashed line and scissors for the exfoliation and cleavage cases illustrate how we cut the crystal.}
   \label{exfbindcleav}
\end{figure}

In the literature, four different energies are used more or less interchangeably when discussing the interlayer binding strength in layered solids. These are the interlayer binding energy, $E_B$, the exfoliation energy, $E_{XF}$, the surface energy $E_{surf}$ and the cleavage energy, $E_{cleav}$. If we, instead of doing full calculations as in the main paper, assume only pairwise interactions between the planes, we get simple expressions that can be compared to see the relations between them. The relation between the two last quantities is simple. The cleavage energy is the energy required to cleave the material in two halves, and the surface energy is the energy required to create one unit of surface by cleavage, and so $E_{surf} = \frac{1}{2}E_{cleav}$. To see how the other quantities are related, we consider the three systems in Figure \ref{exfbindcleav} and assume that we only have pairwise interactions between the planes and ignore all relaxation effects.
The interaction energies between pairs of layers are labelled $\varepsilon_1$, $\varepsilon_2$ and $\varepsilon_3$, for adjacent layers, second nearest layers, third nearest layers, and so on. Ignoring all effects of relaxing the layer positions, which are expected to be small, we see that the exfoliation energy is just the sum, $E_{XF} = \varepsilon_1 + \varepsilon_2 + \ldots = \sum_{n=1}^{\infty} \varepsilon_n$. When stretching the crystal to determine the binding energy we break two of each bonds, but since the bonds are shared we need to divide by two to get, $E_{B} = \frac{1}{2}\sum_{n=1}^{\infty} 2\varepsilon_n = E_{XF}$. The expression for the cleavage energy is  $E_{cleav} = \varepsilon_1 + 2\cdot\varepsilon_2 + 3\cdot\varepsilon_3+\ldots = \sum_{n=1}^{\infty} n \cdot \varepsilon_n$. The $\varepsilon$ interaction energies in a solid will decay more rapidly than the factor $n$ in front of the terms, and if we account for only the first term, all three quantities are equal, and in general we have $E_B = E_{XF} \approx E_{cleav} = 2\cdot E_{surf}$.



\subsection{Dichalcogenide binding energies and exfoliation energies}
We here tabulate the binding energies using RPA (Table \ref{rpaenergies}) as well as the full set of binding energies for the MX$_2$ compounds calculated with the VV10 functional (Table \ref{mx2energies}). Note that the estimate used for the binding energies in the main paper is 0.66 times the binding energies as calculated with VV10. In Subsections A-G we tabulate all data calculated for the comparison of van der Waals functionals. These are the range of experimentally reported values of the $c$ axis lengths, the calculated $c$ axis lengths and the range of their deviations from experiments. We also give the $C_{33}$ lattice constants, the interlayer binding energies, $E_B$ and the bandgap, $E_g$ (the letter ''M'' marks metals).

\clearpage
\begin{table}[htdp]
\caption{Interlayer binding energies ($E_B$), calculated using RPA and NLCF . The NLCF estimates, denoted by an asterisk (*) after the number, were obtained by a VV10 calculation rescaled by a factor 0.66. Two asterisks (**) after the compound name denotes structures not previously reported experimentally.}
\begin{center}
\begin{tabular}{l|c||l|c||l|c}
\multicolumn{2}{c||}{$3d$} & \multicolumn{2}{c||}{$4d$} & \multicolumn{2}{c}{$5d$} \\
Compound & $E_B$ [meV/\AA$^2$] & Compound & $E_B$ [meV/\AA$^2$] & Compound & $E_B$ [meV/\AA$^2$]\\\hline\hline
TiS$_2$ & 18.9 & ZrS$_2$ & 17.0 & HfS$_2$ & 16.1 \\
TiSe$_2$ & 17.4 & ZrSe$_2$ & 18.5 & HfSe$_2$ & 17.1 \\
TiTe$_2$ & 19.7 & ZrTe$_2$ & 16.3 & HfTe$_2$ & 18.7 \\
VS$_2$ & 25.6 & NbS$_2$ & 17.6 & TaS$_2$ & 17.9 \\
VSe$_2$ & 22.3 &NbSe$_2$ & 19.6 & TaSe$_2$ & 19.4 \\
VTe$_2$ & 20.4 &NbTe$_2$ & 23.0 & TaTe$_2$ & 23.2* \\
CrS$_2$ & 19.5* & MoS$_2$ & 20.5 & WS$_2$ & 20.2 \\
CrSe$_2$ & 23.9* & MoSe$_2$ & 19.6 & WSe$_2$ & 20.0 \\
CrTe$_2$** & 29.9* & MoTe$_2$ & 20.8 & WTe$_2$ & 18.5* \\
 & & TcS$_2$ & 19.7* & ReS$_2$ & 19.9* \\
 & & TcSe$_2$** & 18.8* & ReSe$_2$ & 18.2* \\
 & & TcTe$_2$** & 18.8* & ReTe$_2$** & 18.5* \\\hline\hline
CoTe$_2$ & 32.9* & RhTe$_2$ & 36.5* & IrTe$_2$ & 36.3* \\
NiS$_2$** & 24.9* & PdS$_2$ & 21.4* & PtS$_2$ & 20.6 \\
NiSe$_2$** & 29.2* & PdSe$_2$ & 27.8* & PtSe$_2$ & 19.1 \\
NiTe$_2$ & 30.2* & PdTe$_2$ & 40.2 & PtTe$_2$ & 25.1* \\
\end{tabular}
\end{center}
\label{mx2energies}
\end{table}%

\newpage
\subsection{RPA binding energies}
\begin{table}[htdp]
\caption{Binding energies calculated with RPA. The numbering of the compounds in the table is the same as in Figure \ref{rpacomp}.}
\begin{center}
\begin{tabular}{ll|c||llc}
No. & Compound & $E_B$ [meV/\AA$^2$] & No. & Compound & $E_B$ [meV/\AA$^2$]\\\hline\hline
1 & TiS$_2$ & 18.88 & 16 & PdTe$_2$ & 40.17 \\
2 & TiSe$_2$ & 17.39 & 17 & HfS$_2$ & 16.13 \\
3 &  TiTe$_2$ & 19.76 & 18 & HfSe$_2$ & 17.09 \\
4 & VS$_2$ & 25.61 & 19 & HfTe$_2$ & 18.68 \\
5 & VSe$_2$ & 22.26 & 20 & TaS$_2$ & 17.86 \\
6 & VTe$_2$ & 20.39 & 21 & TaSe$_2$ & 19.44 \\
7 & ZrS$_2$ & 16.98 & 22 & WS$_2$ & 20.24 \\
8 & ZrSe$_2$ & 18.53 & 23 & WSe$_2$ & 19.98 \\
9 & ZrTe$_2$ & 16.34 & 24 & PtS$_2$ & 20.55 \\
10 & NbS$_2$ & 17.58 & 25 & PtSe$_2$ & 19.05 \\
11 & NbSe$_2$ & 19.57 & 26 & Graphite\cite{seb} & 18.32 \\
12 & NbTe$_2$ & 23.03 & 27 & BN & 14.49 \\
13 & MoS$_2$ & 20.53 & 28 & PbO & 20.25 \\
14 & MoSe$_2$ & 19.63 & & & \\
15 & MoTe$_2$ & 20.80 & & &
\end{tabular}
\end{center}
\label{rpaenergies}
\end{table}%

\newpage
\subsection{VV10}

\begin{longtable}{l|cccccc}
 Name & $c_{exp}$ [\AA] & $c_{calc}$ & $c$ dev. [\%] & $C_{33}$ [GPa] & $E_B$ [meV/\AA$^2$] & $E_g$ [eV] \\\hline\hline
 \endfirsthead
 
 \endhead
   AgBiP$_2$Se$_6$ & 39.615 & 39.402 & -0.5 & 35.6 &   23.396 &    1.027\\
   BBr$_3$ &  6.847 --  6.864 &  6.754 & -1.4 -- -1.6 & 18.1 &   12.314 &    3.470\\
   BI$_3$ &  7.261 --  7.460 &  7.202 & -0.8 -- -3.5 & 17.9 &   13.956 &    2.471\\
   BN &  6.661 --  6.690 &  6.679 & -0.2 --  0.3 & 41.2 &   25.336 &    4.233\\
   BaFI &  7.962 --  8.102 &  8.053 & -0.6 --  1.1 & 35.8 &   26.387 &    3.843\\
   Bi$_2$Se$_3$ & 28.615 -- 28.636 & 29.014 &  1.3 --  1.4 & 45.9 &   25.596 &    0.249\\
   Bi$_2$Te$_3$ & 30.440 -- 30.497 & 31.061 &  1.9 --  2.0 & 39.8 &   26.193 &    0.734\\
   BiIO &  9.128 --  9.151 &  9.449 &  3.3 --  3.5 & 36.9 &   21.164 &    1.985\\
   C &  6.704 --  6.930 &  6.777 & -0.3 -- -2.2 & 46.1 &   27.073 & M\\
   CdI$_2$ &  6.835 --  6.864 &  6.932 &  1.0 --  1.4 & 28.2 &   16.711 &    2.004\\
   CoTe$_2$ &  5.405 &  5.672 &  4.9 & 55.4 &   49.269 & M\\
   CrSe$_2$ &  5.915 &  6.054 &  2.4 & 38.0 &   30.627 & M\\
   CrSiTe$_3$ & 20.528 -- 20.582 & 20.692 &  0.5 --  0.8 & 34.7 &   24.158 &    0.382\\
   Cu$_2$S &  6.670 --  6.680 &  6.755 &  1.1 --  1.3 & 31.9 &   32.111 & M\\
   GaS & 14.230 -- 15.530 & 15.658 &  0.8 -- 10.0 & 45.6 &   20.144 &    1.800\\
   GaSe & 15.919 -- 15.995 & 16.111 &  0.7 --  1.2 & 40.2 &   20.073 &    1.406\\
   Ge$_2$Sb$_2$Te$_5$ & 16.960 -- 17.239 & 18.452 &  7.0 --  8.8 & 38.6 &   25.880 & M\\
   HfS$_2$ &  5.837 --  5.856 &  5.856 & -0.0 --  0.3 & 46.4 &   24.172 &    0.951\\
   HfSe$_2$ &  6.143 --  6.159 &  6.241 &  1.3 --  1.6 & 47.1 &   25.091 &    0.432\\
   HfTe$_2$ &  6.650 --  6.670 &  6.816 &  2.2 --  2.5 & 41.0 &   27.162 & M\\
   HgI$_2$ & 12.088 -- 12.450 & 12.512 &  0.5 --  3.5 & 22.5 &   17.815 &    0.866\\
   In$_2$Zn$_2$S$_5$ & 46.270 & 46.554 &  0.6 & 93.6 &   47.137 &    0.311\\
   Mg$_2$(P$_2$Se$_6$) & 20.194 & 20.286 &  0.5 & 36.0 &   21.074 &    1.904\\
   MgBr$_2$ &  6.260 --  6.269 &  6.290 &  0.3 --  0.5 & 29.5 &   15.322 &    4.236\\
   MgI$_2$ &  6.862 --  6.895 &  6.875 &  0.2 -- -0.3 & 27.7 &   15.816 &    3.303\\
   MoS$_2$ & 12.290 -- 12.530 & 12.387 &  0.5 -- -1.1 & 61.3 &   30.820 &    0.881\\
   MoS$_2$ & 18.330 -- 18.450 & 18.267 & -0.3 -- -1.0 & 55.2 &   30.514 &    1.182\\
   MoSe$_2$ & 12.900 -- 12.930 & 13.186 &  2.0 --  2.2 & 57.5 &   30.062 &    0.938\\
   MoTe$_2$ & 13.964 -- 13.974 & 14.240 &  1.9 --  2.0 & 50.9 &   30.408 &    0.857\\
   NbS$_2$ & 17.800 -- 17.918 & 18.031 &  0.6 --  1.3 & 65.9 &   30.036 & M\\
   NbSe$_2$ & 12.482 -- 12.550 & 12.743 &  1.5 --  2.1 & 55.0 &   32.888 & M\\
   NbSe$_2$ & 25.230 -- 25.450 & 25.590 &  0.6 --  1.4 & 57.1 &   33.878 & M\\
   NbTe$_2$ &  6.610 &  7.040 &  6.5 & 44.0 &   30.074 & M\\
   Ni$_2$SbTe$_2$ & 15.634 -- 15.682 & 16.177 &  3.2 --  3.5 & 48.8 &   34.636 & M\\
   NiSbSi &  8.179 &  8.224 &  0.6 & 62.4 &   66.807 & M\\
   NiTe$_2$ &  5.251 --  5.308 &  5.556 &  4.7 --  5.8 & 50.4 &   42.807 & M\\
   PbBi$_4$Te$_7$ & 23.600 -- 23.892 & 24.355 &  1.9 --  3.2 & 10.8 &   13.096 &    0.764\\
   PbFI &  8.770 --  8.800 &  9.120 &  3.6 --  4.0 & 26.5 &   19.606 &    1.748\\
   PbO &  4.988 --  5.071 &  5.213 &  2.8 --  4.5 & 25.9 &   26.033 &    1.335\\
   PbSb$_2$Te$_4$ & 41.712 & 41.545 & -0.4 & 5.2 &    9.108 &    0.331\\
   PdTe$_2$ &  5.113 --  5.270 &  5.469 &  3.8 --  7.0 & 72.3 &   46.510 & M\\
   PtS$_2$ &  5.019 --  5.043 &  5.232 &  3.8 --  4.2 & 33.2 &   29.384 &    0.721\\
   PtSe$_2$ &  5.031 --  5.082 &  5.418 &  6.6 --  7.7 & 33.5 &   29.920 & M\\
   PtTe$_2$ &  5.201 --  5.224 &  5.603 &  7.3 --  7.7 & 49.6 &   33.945 & M\\
   Re(AgCl$_3$)$_2$ & 16.731 -- 16.731 & 16.633 & -0.6 -- -0.6 & 39.3 &   23.063 & M\\
   RhTe$_2$ &  5.410 --  5.442 &  5.688 &  4.5 --  5.1 & 59.7 &   51.159 & M\\
   SnS$_2$ &  5.460 --  5.960 &  5.962 &  0.0 --  9.2 & 37.1 &   22.438 &    1.263\\
   SnSe$_2$ &  6.128 --  6.141 &  6.298 &  2.6 --  2.8 & 37.1 &   23.413 &    0.264\\
   SrFI &  8.888 --  8.916 &  8.715 & -2.0 -- -2.3 & 23.8 &   18.975 &    4.117\\
   TaS$_2$ &  5.853 --  5.900 &  5.993 &  1.6 --  2.4 & 57.3 &   30.792 & M\\
   TaS$_2$ & 12.097 -- 12.100 & 12.184 &  0.7 --  0.7 & 57.4 &   31.466 & M\\
   TaSe$_2$ &  6.203 --  6.272 &  6.364 &  1.5 --  2.6 & 56.5 &   31.489 & M\\
   TaSe$_2$ & 12.696 -- 12.720 & 12.828 &  0.8 --  1.0 & 58.3 &   31.999 & M\\
   TaSe$_2$ & 25.143 -- 25.500 & 25.511 &  0.0 --  1.5 & 59.1 &   33.783 & M\\
   Ti$_2$PTe$_2$ & 28.486 & 28.764 &  1.0 & 49.1 &   29.746 & M\\
   TiS$_2$ &  5.680 --  5.716 &  5.752 &  0.6 --  1.3 & 50.6 &   28.618 & M\\
   TiSe$_2$ &  5.981 --  6.011 &  6.112 &  1.7 --  2.2 & 44.2 &   28.223 & M\\
   TiTe$_2$ &  6.459 --  6.539 &  6.712 &  2.6 --  3.9 & 44.9 &   29.462 & M\\
   VBr$_2$ &  6.206 &  6.087 & -1.9 & 34.6 &   19.955 & M\\
   VCl$_2$ &  5.798 --  5.835 &  5.640 & -2.7 -- -3.3 & 34.8 &   19.334 & M\\
   VI$_2$ &  6.714 &  6.582 & -2.0 & 32.0 &   20.622 & M\\
   VS$_2$ &  5.755 --  5.755 &  5.839 &  1.5 --  1.5 & 55.6 &   31.573 & M\\
   VSe$_2$ &  6.048 --  6.150 &  6.296 &  2.4 --  4.1 & 49.0 &   30.349 & M\\
   WS$_2$ & 12.323 -- 12.500 & 12.360 & -0.0 -- -1.1 & 62.1 &   31.020 &    1.134\\
   WS$_2$ & 18.490 & 18.409 & -0.4 & 56.7 &   30.492 &    1.172\\
   WSe$_2$ & 12.960 -- 12.980 & 13.152 &  1.3 --  1.5 & 58.3 &   30.598 &    1.148\\
   Y$_2$I$_2$Ga$_2$ & 11.434 & 11.495 &  0.5 & 44.7 &   19.058 & M\\
   YI$_3$ & 20.880 & 20.901 &  0.1 & 25.2 &   15.495 &    2.407\\
   ZrNCl & 27.178 -- 27.672 & 27.472 &  0.2 --  1.1 & 66.5 &   25.191 &    1.800\\
   ZrS$_2$ &  5.810 --  5.850 &  5.855 &  0.1 --  0.8 & 47.6 &   24.467 &    0.656\\
   ZrSe$_2$ &  6.125 --  6.192 &  6.229 &  0.6 --  1.7 & 47.7 &   25.644 &    0.214\\
   ZrTe$_2$ &  6.630 --  6.660 &  6.805 &  2.2 --  2.6 & 49.9 &   28.983 & M\\
\end{longtable}

\newpage
\subsection{vdW-DF1}
\begin{longtable}{l|cccccc}
 Name & $c_{exp}$ [\AA] & $c_{calc}$ & $c$ dev. [\%] & $C_{33}$ [GPa] & $E_B$ [meV/\AA$^2$] & $E_g$ [eV] \\\hline\hline
   AgBiP$_2$Se$_6$ & 39.615 & 40.069 &  1.1 & 13.8 &   13.192 &    1.306\\
   BBr$_3$ &  6.847 --  6.864 &  7.247 &  5.6 --  5.8 & 8.8 &   11.046 &    3.744\\
   BI$_3$ &  7.261 --  7.460 &  7.671 &  2.8 --  5.6 & 8.1 &   10.619 &    2.636\\
   BN &  6.661 --  6.690 &  6.926 &  3.5 --  4.0 & 20.2 &   18.093 &    4.649\\
   BaFI &  7.962 --  8.102 &  8.986 & 10.9 -- 12.9 & 10.4 &   13.801 &    4.317\\
   Bi$_2$Se$_3$ & 28.615 -- 28.636 & 29.741 &  3.9 --  3.9 & 17.2 &   12.914 &    0.988\\
   Bi$_2$Te$_3$ & 30.440 -- 30.497 & 31.907 &  4.6 --  4.8 & 14.5 &   12.159 &    1.150\\
   BiIO &  9.128 --  9.151 & 10.202 & 11.5 -- 11.8 & 19.4 &   12.800 &    2.115\\
   C &  6.704 --  6.930 &  7.199 &  3.9 --  7.4 & 23.0 &   18.900 & M\\
   CdI$_2$ &  6.835 --  6.864 &  7.572 & 10.3 -- 10.8 & 15.0 &   11.418 &    2.145\\
   CoTe$_2$ &  5.405 &  6.043 & 11.8 & 27.1 &   25.531 & M\\
   CrSe$_2$ &  5.915 &  6.697 & 13.2 & 18.4 &   15.704 & M\\
   CrSiTe$_3$ & 20.528 -- 20.582 & 21.193 &  3.0 --  3.2 & 14.8 &   13.101 &    0.366\\
   Cu$_2$S &  6.670 --  6.680 &  7.250 &  8.5 --  8.7 & 11.0 &   16.494 & M\\
   GaS & 14.230 -- 15.530 & 16.152 &  4.0 -- 13.5 & 23.8 &   14.063 &    2.339\\
   GaSe & 15.919 -- 15.995 & 16.759 &  4.8 --  5.3 & 19.7 &   12.731 &    1.929\\
   Ge$_2$Sb$_2$Te$_5$ & 16.960 -- 17.239 & 20.898 & 21.2 -- 23.2 & 24.2 &   14.745 &    0.314\\
   HfS$_2$ &  5.837 --  5.856 &  6.354 &  8.5 --  8.9 & 21.8 &   15.686 &    1.318\\
   HfSe$_2$ &  6.143 --  6.159 &  6.777 & 10.0 -- 10.3 & 19.4 &   14.739 &    0.598\\
   HfTe$_2$ &  6.650 --  6.670 &  7.439 & 11.5 -- 11.9 & 16.9 &   14.293 & M\\
   HgI$_2$ & 12.088 -- 12.450 & 13.357 &  7.3 -- 10.5 & 10.5 &   11.229 &    1.207\\
   In$_2$Zn$_2$S$_5$ & 46.270 & 47.102 &  1.8 & 53.3 &   31.454 &    0.321\\
   Mg$_2$(P$_2$Se$_6$) & 20.194 & 20.882 &  3.4 & 18.2 &   13.381 &    2.031\\
   MgBr$_2$ &  6.260 --  6.269 &  6.719 &  7.2 --  7.3 & 18.9 &   13.045 &    4.811\\
   MgI$_2$ &  6.862 --  6.895 &  7.399 &  7.3 --  7.8 & 15.7 &   11.746 &    3.631\\
   MoS$_2$ & 12.290 -- 12.530 & 12.812 &  2.2 --  4.2 & 24.4 &   16.707 &    1.396\\
   MoSe$_2$ & 12.900 -- 12.930 & 13.662 &  5.7 --  5.9 & 20.4 &   15.329 &    1.252\\
   MoTe$_2$ & 13.964 -- 13.974 & 14.683 &  5.1 --  5.2 & 17.6 &   14.368 &    0.958\\
   NbS$_2$ & 17.800 -- 17.918 & 18.346 &  2.4 --  3.1 & 26.8 &   16.662 & M\\
   NbSe$_2$ & 12.482 -- 12.550 & 13.278 &  5.8 --  6.4 & 21.8 &   16.750 & M\\
   NbSe$_2$ & 25.230 -- 25.450 & 26.127 &  2.7 --  3.6 & 22.0 &   17.523 & M\\
   NbTe$_2$ &  6.610 &  7.710 & 16.6 & 15.8 &   14.637 & M\\
   Ni$_2$SbTe$_2$ & 15.634 -- 15.682 & 16.978 &  8.3 --  8.6 & 16.1 &   15.047 & M\\
   NiSbSi &  8.179 &  8.578 &  4.9 & 53.6 &   38.880 & M\\
   NiTe$_2$ &  5.251 --  5.308 &  5.973 & 12.5 -- 13.8 & 19.1 &   19.249 & M\\
   PbBi$_4$Te$_7$ & 23.600 -- 23.892 & 25.437 &  6.5 --  7.8 & 4.0 &    5.939 &    1.143\\
   PbFI &  8.770 --  8.800 & 10.115 & 14.9 -- 15.3 & 15.6 &   12.056 &    2.381\\
   PbO &  4.988 --  5.071 &  6.084 & 20.0 -- 22.0 & 7.8 &    9.923 &    2.215\\
   PbSb$_2$Te$_4$ & 41.712 & 42.920 &  2.9 & 2.3 &    4.072 &    0.821\\
   PdTe$_2$ &  5.113 --  5.270 &  5.699 &  8.1 -- 11.5 & 35.6 &   19.653 & M\\
   PtS$_2$ &  5.019 --  5.043 &  6.013 & 19.2 -- 19.8 & 16.3 &   14.851 &    1.264\\
   PtSe$_2$ &  5.031 --  5.082 &  6.312 & 24.2 -- 25.5 & 12.6 &   13.449 &    0.831\\
   PtTe$_2$ &  5.201 --  5.224 &  6.782 & 29.8 -- 30.4 & 5.9 &   12.019 &    0.199\\
   Re(AgCl$_3$)$_2$ & 16.731 -- 16.731 & 17.198 &  2.8 --  2.8 & 31.9 &   18.445 & M\\
   RhTe$_2$ &  5.410 --  5.442 &  6.013 & 10.5 -- 11.1 & 29.5 &   25.776 & M\\
   SnS$_2$ &  5.460 --  5.960 &  6.524 &  9.5 -- 19.5 & 19.2 &   14.396 &    1.445\\
   SnSe$_2$ &  6.128 --  6.141 &  6.930 & 12.8 -- 13.1 & 16.0 &   13.348 &    0.505\\
   SrFI &  8.888 --  8.916 &  9.766 &  9.5 --  9.9 & 13.9 &   11.802 &    4.379\\
   TaS$_2$ &  5.853 --  5.900 &  6.512 & 10.4 -- 11.3 & 25.4 &   17.714 & M\\
   TaS$_2$ & 12.097 -- 12.100 & 12.732 &  5.2 --  5.3 & 24.4 &   17.532 & M\\
   TaSe$_2$ &  6.203 --  6.272 &  6.880 &  9.7 -- 10.9 & 22.8 &   16.798 & M\\
   TaSe$_2$ & 12.696 -- 12.720 & 13.331 &  4.8 --  5.0 & 21.1 &   16.339 & M\\
   TaSe$_2$ & 25.143 -- 25.500 & 25.786 &  1.1 --  2.6 & 22.5 &   17.380 & M\\
   Ti$_2$PTe$_2$ & 28.486 & 29.620 &  4.0 & 23.0 &   14.255 & M\\
   TiS$_2$ &  5.680 --  5.716 &  6.301 & 10.2 -- 10.9 & 22.6 &   17.093 & M\\
   TiSe$_2$ &  5.981 --  6.011 &  6.723 & 11.8 -- 12.4 & 19.7 &   15.715 & M\\
   TiTe$_2$ &  6.459 --  6.539 &  7.429 & 13.6 -- 15.0 & 16.9 &   14.876 & M\\
   TlCrTe$_2$ &  7.839 --  7.935 &  8.345 &  5.2 --  6.5 & 42.7 &   48.379 & M\\
   VBr$_2$ &  6.206 &  6.612 &  6.5 & 20.1 &   14.183 & M\\
   VCl$_2$ &  5.798 --  5.835 &  6.137 &  5.2 --  5.9 & 22.3 &   15.206 & M\\
   VI$_2$ &  6.714 &  7.202 &  7.3 & 16.4 &   12.861 & M\\
   VS$_2$ &  5.755 --  5.755 &  6.392 & 11.1 -- 11.1 & 24.8 &   18.121 & M\\
   VSe$_2$ &  6.048 --  6.150 &  6.861 & 11.6 -- 13.4 & 21.4 &   16.334 & M\\
   WS$_2$ & 12.323 -- 12.500 & 12.874 &  3.0 --  4.5 & 24.3 &   16.575 &    1.572\\
   WS$_2$ & 18.490 & 18.983 &  2.7 & 23.5 &   16.399 &    1.582\\
   WSe$_2$ & 12.960 -- 12.980 & 13.707 &  5.6 --  5.8 & 21.0 &   15.203 &    1.342\\
   Y$_2$I$_2$Ga$_2$ & 11.434 & 12.196 &  6.7 & 23.8 &   12.040 & M\\
   YI$_3$ & 20.880 & 21.511 &  3.0 & 13.3 &    9.864* &    2.736\\
   ZrNCl & 27.178 -- 27.672 & 28.067 &  1.4 --  3.3 & 38.1 &   16.087 &    1.907\\
   ZrS$_2$ &  5.810 --  5.850 &  6.358 &  8.7 --  9.4 & 21.4 &   15.775 &    1.050\\
   ZrSe$_2$ &  6.125 --  6.192 &  6.771 &  9.3 -- 10.5 & 19.2 &   15.017 &    0.363\\
   ZrTe$_2$ &  6.630 --  6.660 &  7.446 & 11.8 -- 12.3 & 17.2 &   14.856 & M\\
\end{longtable}

\newpage
\subsection{vdW-DF2}
\begin{longtable}{l|cccccc}
 Name & $c_{exp}$ [\AA] & $c_{calc}$ & $c$ dev. [\%] & $C_{33}$ [GPa] & $E_B$ [meV/\AA$^2$] & $E_g$ [eV] \\\hline\hline
   AgBiP$_2$Se$_6$ & 39.615 & 40.016 &  1.0 & 23.0 &   14.220 &    0.987\\
   BBr$_3$ &  6.847 --  6.864 &  7.011 &  2.1 --  2.4 & 13.1 &   10.020 &    3.653\\
   BI$_3$ &  7.261 --  7.460 &  7.445 & -0.2 --  2.5 & 11.7 &   10.244 &    2.569\\
   BN &  6.661 --  6.690 &  6.878 &  2.8 --  3.3 & 29.7 &   17.677 &    4.567\\
   BaFI &  7.962 --  8.102 &  8.705 &  7.4 --  9.3 & 19.0 &   15.047 &    4.078\\
   Bi$_2$Se$_3$ & 28.615 -- 28.636 & 29.750 &  3.9 --  4.0 & 27.2 &   14.245 &    0.695\\
   Bi$_2$Te$_3$ & 30.440 -- 30.497 & 31.750 &  4.1 --  4.3 & 22.8 &   13.570 &    0.873\\
   BiIO &  9.128 --  9.151 & 10.124 & 10.6 -- 10.9 & 27.5 &   12.775 &    1.968\\
   C &  6.704 --  6.930 &  6.944 &  0.2 --  3.6 & 33.5 &   18.630 & M\\
   CdI$_2$ &  6.835 --  6.864 &  7.411 &  8.0 --  8.4 & 21.3 &   11.056 &    1.877\\
   CoTe$_2$ &  5.405 &  6.249 & 15.6 & 31.8 &   24.201 & M\\
   CrSe$_2$ &  5.915 &  6.554 & 10.8 & 31.8 &   17.177 & M\\
   CrSiTe$_3$ & 20.528 -- 20.582 & 21.282 &  3.4 --  3.7 & 22.7 &   13.863 &    0.323\\
   Cu$_2$S &  6.670 --  6.680 &  7.156 &  7.1 --  7.3 & 16.9 &   17.571 & M\\
   Fe(PSe$_3$) & 19.800 -- 19.812 & 20.105 &  1.5 --  1.5 & 25.0 &   14.449 & M\\
   GaS & 14.230 -- 15.530 & 15.838 &  2.0 -- 11.3 & 33.6 &   13.645 &    2.163\\
   GaSe & 15.919 -- 15.995 & 16.463 &  2.9 --  3.4 & 26.7 &   12.674 &    1.691\\
   Ge$_2$Sb$_2$Te$_5$ & 16.960 -- 17.239 & 20.771 & 20.5 -- 22.5 & 17.8 &   12.524 &    0.105\\
   HfS$_2$ &  5.837 --  5.856 &  6.162 &  5.2 --  5.6 & 32.7 &   16.097 &    1.245\\
   HfSe$_2$ &  6.143 --  6.159 &  6.604 &  7.2 --  7.5 & 30.7 &   15.717 &    0.486\\
   HfTe$_2$ &  6.650 --  6.670 &  7.320 &  9.7 -- 10.1 & 27.6 &   15.682 & M\\
   HgI$_2$ & 12.088 -- 12.450 & 13.145 &  5.6 --  8.8 & 15.4 &   11.594 &    0.867\\
   In$_2$Zn$_2$S$_5$ & 46.270 & 47.249 &  2.1 & 65.3 &   31.457 &    0.315\\
   Mg$_2$(P$_2$Se$_6$) & 20.194 & 20.386 &  1.0 & 25.7 &   13.675 &    1.801\\
   MgBr$_2$ &  6.260 --  6.269 &  6.532 &  4.2 --  4.3 & 24.8 &   11.814 &    4.481\\
   MgI$_2$ &  6.862 --  6.895 &  7.226 &  4.8 --  5.3 & 22.1 &   11.314 &    3.347\\
   MoS$_2$ & 12.290 -- 12.530 & 12.682 &  1.2 --  3.2 & 38.6 &   17.870 &    1.193\\
   MoS$_2$ & 18.330 -- 18.450 & 18.798 &  1.9 --  2.6 & 39.1 &   17.641 &    1.384\\
   MoSe$_2$ & 12.900 -- 12.930 & 13.401 &  3.6 --  3.9 & 34.3 &   16.637 &    1.098\\
   MoTe$_2$ & 13.964 -- 13.974 & 14.762 &  5.6 --  5.7 & 28.6 &   16.027 &    0.830\\
   NbS$_2$ & 17.800 -- 17.918 & 18.324 &  2.3 --  2.9 & 45.4 &   20.100 & M\\
   NbSe$_2$ & 12.482 -- 12.550 & 13.218 &  5.3 --  5.9 & 37.2 &   18.768 & M\\
   NbSe$_2$ & 25.230 -- 25.450 & 25.680 &  0.9 --  1.8 & 36.8 &   19.623 & M\\
   NbTe$_2$ &  6.610 &  7.643 & 15.6 & 25.8 &   15.819 & M\\
   Ni$_2$SbTe$_2$ & 15.634 -- 15.682 & 16.923 &  7.9 --  8.2 & 26.0 &   17.182 & M\\
   NiTe$_2$ &  5.251 --  5.308 &  6.226 & 17.3 -- 18.6 & 25.7 &   19.479 & M\\
   PbBi$_4$Te$_7$ & 23.600 -- 23.892 & 25.971 &  8.7 -- 10.0 & 6.8 &    6.612 &    0.901\\
   PbFI &  8.770 --  8.800 &  9.938 & 12.9 -- 13.3 & 21.4 &   11.807 &    2.094\\
   PbO &  4.988 --  5.071 &  5.904 & 16.4 -- 18.4 & 12.7 &   11.012 &    1.910\\
   PbSb$_2$Te$_4$ & 41.712 & 43.940 &  5.3 & 2.9 &    4.565 &    0.499\\
   PdTe$_2$ &  5.113 --  5.270 &  6.045 & 14.7 -- 18.2 & 23.7 &   20.815 & M\\
   PtS$_2$ &  5.019 --  5.043 &  5.826 & 15.5 -- 16.1 & 23.7 &   15.883 &    1.027\\
   PtSe$_2$ &  5.031 --  5.082 &  6.146 & 20.9 -- 22.2 & 20.2 &   15.055 &    0.563\\
   PtTe$_2$ &  5.201 --  5.224 &  6.523 & 24.9 -- 25.4 & 10.1 &   13.479 & M\\
   RhTe$_2$ &  5.410 --  5.442 &  6.287 & 15.5 -- 16.2 & 29.5 &   24.516 & M\\
   SnS$_2$ &  5.460 --  5.960 &  6.326 &  6.1 -- 15.9 & 27.2 &   14.528 &    1.220\\
   SnSe$_2$ &  6.128 --  6.141 &  6.752 & 10.0 -- 10.2 & 24.4 &   14.192 &    0.220\\
   SrFI &  8.888 --  8.916 &  9.484 &  6.4 --  6.7 & 18.8 &   11.620 &    4.029\\
   TaS$_2$ &  5.853 --  5.900 &  6.333 &  7.3 --  8.2 & 39.6 &   19.324 & M\\
   TaS$_2$ & 12.097 -- 12.100 & 12.567 &  3.9 --  3.9 & 39.0 &   18.879 & M\\
   TaSe$_2$ &  6.203 --  6.272 &  6.752 &  7.7 --  8.9 & 36.7 &   18.874 & M\\
   TaSe$_2$ & 12.696 -- 12.720 & 13.238 &  4.1 --  4.3 & 35.3 &   18.289 & M\\
   TaSe$_2$ & 25.143 -- 25.500 & 25.899 &  1.6 --  3.0 & 36.9 &   19.645 & M\\
   Ti$_2$PTe$_2$ & 28.486 & 29.477 &  3.5 & 35.0 &   15.690 & M\\
   TiS$_2$ &  5.680 --  5.716 &  6.099 &  6.7 --  7.4 & 35.7 &   18.170 & M\\
   TiSe$_2$ &  5.981 --  6.011 &  6.559 &  9.1 --  9.7 & 31.8 &   17.106 & M\\
   TiTe$_2$ &  6.459 --  6.539 &  7.334 & 12.2 -- 13.5 & 26.1 &   16.115 & M\\
   VBr$_2$ &  6.206 &  6.458 &  4.1 & 27.1 &   13.669 & M\\
   VCl$_2$ &  5.798 --  5.835 &  5.953 &  2.0 --  2.7 & 30.1 &   13.961 & M\\
   VI$_2$ &  6.714 &  7.079 &  5.4 & 22.9 &   13.022 & M\\
   VS$_2$ &  5.755 --  5.755 &  6.214 &  8.0 --  8.0 & 38.1 &   19.617 & M\\
   VSe$_2$ &  6.048 --  6.150 &  6.731 &  9.5 -- 11.3 & 32.7 &   17.862 & M\\
   WS$_2$ & 12.323 -- 12.500 & 12.742 &  1.9 --  3.4 & 38.8 &   17.901 &    1.367\\
   WS$_2$ & 18.490 & 18.883 &  2.1 & 36.1 &   17.552 &    1.391\\
   WSe$_2$ & 12.960 -- 12.980 & 13.452 &  3.6 --  3.8 & 34.3 &   16.601 &    1.189\\
   Y$_2$I$_2$Ga$_2$ & 11.434 & 12.105 &  5.9 & 32.1 &   11.944 &    0.202\\
   YI$_3$ & 20.880 & 21.316 &  2.1 & 18.3 &   10.795 &    2.540\\
   ZrNCl & 27.178 -- 27.672 & 27.712 &  0.1 --  2.0 & 51.2 &   15.125 &    1.894\\
   ZrS$_2$ &  5.810 --  5.850 &  6.165 &  5.4 --  6.1 & 32.8 &   16.316 &    0.948\\
   ZrSe$_2$ &  6.125 --  6.192 &  6.602 &  6.6 --  7.8 & 31.1 &   16.194 &    0.276\\
   ZrTe$_2$ &  6.630 --  6.660 &  7.350 & 10.4 -- 10.9 & 28.0 &   16.391 & M\\
\end{longtable}

\newpage
\subsection{vdW-DF1 (PBE)}
\begin{longtable}{l|cccccc}
 Name & $c_{exp}$ [\AA] & $c_{calc}$ & $c$ dev. [\%] & $C_{33}$ [GPa] & $E_B$ [meV/\AA$^2$] & $E_g$ [eV] \\\hline\hline
   AgBiP$_2$Se$_6$ & 39.615 & 40.444 &  2.1 & 24.7 &   19.024 &    1.194\\
   BBr$_3$ &  6.847 --  6.864 &  6.915 &  0.7 --  1.0 & 14.6 &   15.120 &    3.689\\
   BI$_3$ &  7.261 --  7.460 &  7.396 &  0.4 --  1.9 & 12.7 &   14.808 &    2.615\\
   BN &  6.661 --  6.690 &  6.696 &  0.1 --  0.5 & 32.4 &   24.689 &    4.481\\
   BaFI &  7.962 --  8.102 &  8.262 &  2.0 --  3.8 & 27.9 &   22.804 &    4.089\\
   Bi$_2$Se$_3$ & 28.615 -- 28.636 & 29.325 &  2.4 --  2.5 & 28.7 &   19.311 &    0.596\\
   Bi$_2$Te$_3$ & 30.440 -- 30.497 & 31.290 &  2.6 --  2.8 & 24.0 &   18.530 &    0.978\\
   BiIO &  9.128 --  9.151 &  9.722 &  6.2 --  6.5 & 30.0 &   17.876 &    2.086\\
   C &  6.704 --  6.930 &  6.892 & -0.6 --  2.8 & 35.6 &   25.671 & M\\
   CdI$_2$ &  6.835 --  6.864 &  7.170 &  4.5 --  4.9 & 22.1 &   15.682 &    2.124\\
   CoTe$_2$ &  5.405 &  5.790 &  7.1 & 41.0 &   37.809 & M\\
   CrSe$_2$ &  5.915 &  6.277 &  6.1 & 32.9 &   23.394 & M\\
   Cu$_2$S &  6.670 --  6.680 &  6.804 &  1.9 --  2.0 & 20.6 &   24.635 & M\\
   Fe(PSe$_3$) & 19.800 -- 19.812 & 19.983 &  0.9 --  0.9 & 27.1 &   19.611 & M\\
   GaS & 14.230 -- 15.530 & 15.688 &  1.0 -- 10.2 & 35.4 &   19.190 &    2.139\\
   GaSe & 15.919 -- 15.995 & 16.403 &  2.6 --  3.0 & 29.2 &   17.673 &    1.617\\
   Ge$_2$Sb$_2$Te$_5$ & 16.960 -- 17.239 & 19.099 & 10.8 -- 12.6 & 26.6 &   18.182 &    0.114\\
   HfS$_2$ &  5.837 --  5.856 &  6.039 &  3.1 --  3.5 & 33.9 &   21.470 &    1.250\\
   HfSe$_2$ &  6.143 --  6.159 &  6.431 &  4.4 --  4.7 & 31.8 &   20.930 &    0.554\\
   HfTe$_2$ &  6.650 --  6.670 &  7.062 &  5.9 --  6.2 & 28.3 &   20.740 & M\\
   HgI$_2$ & 12.088 -- 12.450 & 12.914 &  3.7 --  6.8 & 16.5 &   16.443 &    1.033\\
   In$_2$Zn$_2$S$_5$ & 46.270 & 46.604 &  0.7 & 76.9 &   41.288 &    0.151\\
   Mg$_2$(P$_2$Se$_6$) & 20.194 & 20.456 &  1.3 & 27.2 &   18.637 &    1.994\\
   MgBr$_2$ &  6.260 --  6.269 &  6.397 &  2.0 --  2.2 & 27.0 &   17.262 &    4.667\\
   MgI$_2$ &  6.862 --  6.895 &  7.049 &  2.2 --  2.7 & 23.1 &   15.808 &    3.544\\
   MoS$_2$ & 12.290 -- 12.530 & 12.586 &  0.4 --  2.4 & 39.3 &   23.612 &    1.105\\
   MoS$_2$ & 18.330 -- 18.450 & 18.716 &  1.4 --  2.1 & 36.4 &   23.272 &    1.309\\
   MoSe$_2$ & 12.900 -- 12.930 & 13.385 &  3.5 --  3.8 & 35.6 &   21.895 &    1.137\\
   MoTe$_2$ & 13.964 -- 13.974 & 14.372 &  2.9 --  2.9 & 31.0 &   20.880 &    0.932\\
   NbS$_2$ & 17.800 -- 17.918 & 18.315 &  2.2 --  2.9 & 45.3 &   23.840 & M\\
   NbSe$_2$ & 12.482 -- 12.550 & 12.979 &  3.4 --  4.0 & 34.6 &   24.501 & M\\
   NbSe$_2$ & 25.230 -- 25.450 & 25.390 & -0.2 --  0.6 & 39.2 &   25.553 & M\\
   NbTe$_2$ &  6.610 &  7.278 & 10.1 & 26.6 &   21.807 & M\\
   Ni$_2$SbTe$_2$ & 15.634 -- 15.682 & 16.241 &  3.6 --  3.9 & 29.6 &   23.715 & M\\
   NiSbSi &  8.179 &  8.368 &  2.3 & 57.2 &   52.245 & M\\
   NiTe$_2$ &  5.251 --  5.308 &  5.659 &  6.6 --  7.8 & 35.6 &   31.309 & M\\
   PbBi$_4$Te$_7$ & 23.600 -- 23.892 & 24.683 &  3.3 --  4.6 & 6.7 &    9.324 &    0.972\\
   PbFI &  8.770 --  8.800 &  9.491 &  7.9 --  8.2 & 20.9 &   17.151 &    2.099\\
   PbO &  4.988 --  5.071 &  5.546 &  9.4 -- 11.2 & 15.4 &   17.128 &    1.763\\
   PbSb$_2$Te$_4$ & 41.712 & 43.166 &  3.5 & 3.3 &    6.300 &    0.576\\
   PdTe$_2$ &  5.113 --  5.270 &  5.522 &  4.8 --  8.0 & 55.7 &   33.171 & M\\
   PtS$_2$ &  5.019 --  5.043 &  5.562 & 10.3 -- 10.8 & 23.0 &   22.131 &    1.004\\
   PtSe$_2$ &  5.031 --  5.082 &  5.786 & 13.9 -- 15.0 & 18.3 &   20.978 &    0.364\\
   PtTe$_2$ &  5.201 --  5.224 &  5.744 & 10.0 -- 10.4 & 28.3 &   22.009 & M\\
   Re(AgCl$_3$)$_2$ & 16.731 -- 16.731 & 16.767 &  0.2 --  0.2 & 34.6 &   23.806 & M\\
   RhTe$_2$ &  5.410 --  5.442 &  5.781 &  6.2 --  6.9 & 45.5 &   39.102 & M\\
   SnS$_2$ &  5.460 --  5.960 &  6.180 &  3.7 -- 13.2 & 27.9 &   20.046 &    1.368\\
   SnSe$_2$ &  6.128 --  6.141 &  6.538 &  6.5 --  6.7 & 25.0 &   19.214 &    0.431\\
   SrFI &  8.888 --  8.916 &  9.103 &  2.1 --  2.4 & 17.1 &   17.039 &    4.340\\
   TaS$_2$ &  5.853 --  5.900 &  6.188 &  4.9 --  5.7 & 39.5 &   25.120 & M\\
   TaS$_2$ & 12.097 -- 12.100 & 12.285 &  1.5 --  1.6 & 39.1 &   24.730 & M\\
   TaSe$_2$ &  6.203 --  6.272 &  6.546 &  4.4 --  5.5 & 38.0 &   24.250 & M\\
   TaSe$_2$ & 12.696 -- 12.720 & 13.034 &  2.5 --  2.7 & 36.4 &   23.756 & M\\
   TaSe$_2$ & 25.143 -- 25.500 & 25.664 &  0.6 --  2.1 & 38.2 &   25.247 & M\\
   Ti$_2$PTe$_2$ & 28.486 & 29.157 &  2.4 & 36.1 &   20.769 & M\\
   TiS$_2$ &  5.680 --  5.716 &  5.950 &  4.1 --  4.8 & 36.1 &   24.552 & M\\
   TiSe$_2$ &  5.981 --  6.011 &  6.339 &  5.5 --  6.0 & 29.6 &   22.952 & M\\
   TiTe$_2$ &  6.459 --  6.539 &  6.984 &  6.8 --  8.1 & 21.2 &   22.016 & M\\
   TlCrTe$_2$ &  7.839 --  7.935 &  8.062 &  1.6 --  2.8 & 52.3 &   59.521 & M\\
   VBr$_2$ &  6.206 &  6.252 &  0.7 & 28.7 &   19.446 & M\\
   VCl$_2$ &  5.798 --  5.835 &  5.795 & -0.0 -- -0.7 & 31.2 &   20.499 & M\\
   VI$_2$ &  6.714 &  6.809 &  1.4 & 23.9 &   18.024 & M\\
   VS$_2$ &  5.755 --  5.755 &  6.048 &  5.1 --  5.1 & 38.3 &   26.047 & M\\
   VSe$_2$ &  6.048 --  6.150 &  6.496 &  5.6 --  7.4 & 33.9 &   23.787 & M\\
   WS$_2$ & 12.323 -- 12.500 & 12.652 &  1.2 --  2.7 & 39.9 &   23.341 &    1.282\\
   WS$_2$ & 18.490 & 18.833 &  1.9 & 37.2 &   23.001 &    1.310\\
   WSe$_2$ & 12.960 -- 12.980 & 13.442 &  3.6 --  3.7 & 36.5 &   21.704 &    1.279\\
   Y$_2$I$_2$Ga$_2$ & 11.434 & 11.743 &  2.7 & 33.6 &   16.632 & M\\
   YI$_3$ & 20.880 & 20.944 &  0.3 & 21.3 &   15.188 &    2.697\\
   ZrNCl & 27.178 -- 27.672 & 27.466 &  0.2 --  1.1 & 55.2 &   21.413 &    1.876\\
   ZrS$_2$ &  5.810 --  5.850 &  6.038 &  3.2 --  3.9 & 34.0 &   21.840 &    0.966\\
   ZrSe$_2$ &  6.125 --  6.192 &  6.421 &  3.7 --  4.8 & 32.5 &   21.522 &    0.325\\
   ZrTe$_2$ &  6.630 --  6.660 &  7.045 &  5.8 --  6.3 & 24.3 &   21.868 & M\\
\end{longtable}

\newpage
\subsection{LDA}
\begin{longtable}{l|cccccc}
 Name & $c_{exp}$ [\AA] & $c_{calc}$ & $c$ dev. [\%] & $C_{33}$ [GPa] & $E_B$ [meV/\AA$^2$] & $E_g$ [eV] \\\hline\hline
   AgBiP$_2$Se$_6$ & 39.615 & 39.390 & -0.6 & 33.0 &   11.962 &    1.050\\
   BBr$_3$ &  6.847 --  6.864 &  6.508 & -4.9 -- -5.2 & 21.1 &    8.090 &    3.597\\
   BI$_3$ &  7.261 --  7.460 &  6.977 & -3.9 -- -6.5 & 15.3 &    8.810 &    2.453\\
   BN &  6.661 --  6.690 &  6.579 & -1.2 -- -1.7 & 28.5 &   10.191 &    4.038\\
   BaFI &  7.962 --  8.102 &  7.470 & -6.2 -- -7.8 & 53.3 &   22.712 &    3.865\\
   BiIO &  9.128 --  9.151 &  8.873 & -2.8 -- -3.0 & 31.5 &   10.163 &    1.830\\
   C &  6.704 --  6.930 &  6.750 &  0.7 -- -2.6 & 29.9 &    9.432 & M\\
   CdI$_2$ &  6.835 --  6.864 &  6.551 & -4.1 -- -4.6 & 23.5 &    7.926 &    2.171\\
   CoTe$_2$ &  5.405 &  5.172 & -4.3 & 72.6 &   43.277 & M\\
   CrSe$_2$ &  5.915 &  5.476 & -7.4 & 72.4 &   24.679 & M\\
   CrSiTe$_3$ & 20.528 -- 20.582 & 20.135 & -1.9 -- -2.2 & 26.5 &   13.402 & M\\
   Cu$_2$S &  6.670 --  6.680 &  6.275 & -5.9 -- -6.1 & 40.9 &   23.929 & M\\
   Fe(PSe$_3$) & 19.800 -- 19.812 & 18.548 & -6.3 -- -6.4 & 72.2 &   25.699 &    0.184\\
   GaS & 14.230 -- 15.530 & 15.201 & -0.4 --  6.8 & 36.7 &    8.807 &    1.624\\
   GaSe & 15.919 -- 15.995 & 15.636 & -1.8 -- -2.2 & 36.9 &    9.120 &    1.253\\
   Ge$_2$Sb$_2$Te$_5$ & 16.960 -- 17.239 & 16.797 & -1.0 -- -2.6 & 70.5 &   20.529 & M\\
   HfS$_2$ &  5.837 --  5.856 &  5.630 & -3.5 -- -3.9 & 39.7 &   11.684 &    0.797\\
   HfSe$_2$ &  6.143 --  6.159 &  5.939 & -3.3 -- -3.6 & 42.2 &   13.024 &    0.210\\
   HfTe$_2$ &  6.650 --  6.670 &  6.397 & -3.8 -- -4.1 & 53.2 &   17.611 & M\\
   HgI$_2$ & 12.088 -- 12.450 & 12.084 & -0.0 -- -2.9 & 21.0 &   11.588 &    0.918\\
   In$_2$Zn$_2$S$_5$ & 46.270 & 45.702 & -1.2 & 103.4 &   41.971 &    0.295\\
   Mg$_2$(P$_2$Se$_6$) & 20.194 & 19.992 & -1.0 & 30.9 &    9.809 &    1.956\\
   MgBr$_2$ &  6.260 --  6.269 &  6.061 & -3.2 -- -3.3 & 21.9 &    7.004 &    4.151\\
   MgI$_2$ &  6.862 --  6.895 &  6.596 & -3.9 -- -4.3 & 22.9 &    7.188 &    3.144\\
   MoS$_2$ & 12.290 -- 12.530 & 12.135 & -1.3 -- -3.2 & 53.0 &   13.412 &    0.727\\
   MoS$_2$ & 18.330 -- 18.450 & 18.295 & -0.2 -- -0.8 & 47.8 &   13.506 &    0.891\\
   MoSe$_2$ & 12.900 -- 12.930 & 12.781 & -0.9 -- -1.2 & 53.8 &   13.924 &    0.832\\
   MoTe$_2$ & 13.964 -- 13.974 & 13.777 & -1.3 -- -1.4 & 55.0 &   15.468 &    0.796\\
   NbS$_2$ & 17.800 -- 17.918 & 17.898 & -0.1 --  0.6 & 57.4 &   14.623 & M\\
   NbSe$_2$ & 12.482 -- 12.550 & 12.312 & -1.4 -- -1.9 & 74.4 &   22.091 & M\\
   NbSe$_2$ & 25.230 -- 25.450 & 25.117 & -0.4 -- -1.3 & 74.6 &   22.524 & M\\
   NbTe$_2$ &  6.610 &  6.475 & -2.0 & 61.6 &   22.503 & M\\
   Ni$_2$SbTe$_2$ & 15.634 -- 15.682 & 15.316 & -2.0 -- -2.3 & 86.2 &   30.896 & M\\
   NiSbSi &  8.179 &  7.266 & -11.2 & 107.2 &   41.165 & M\\
   NiTe$_2$ &  5.251 --  5.308 &  5.038 & -4.1 -- -5.1 & 86.6 &   44.862 & M\\
   PbBi$_4$Te$_7$ & 23.600 -- 23.892 & 23.740 &  0.6 -- -0.6 & 14.0 &    9.721 &    0.506\\
   PbFI &  8.770 --  8.800 &  8.196 & -6.5 -- -6.9 & 21.9 &   12.007 &    1.434\\
   PbO &  4.988 --  5.071 &  4.847 & -2.8 -- -4.4 & 24.9 &   19.082 &    1.405\\
   PbSb$_2$Te$_4$ & 41.712 & 41.354 & -0.9 & 7.6 &    7.056 &    0.414\\
   PdTe$_2$ &  5.113 --  5.270 &  5.018 & -1.9 -- -4.8 & 94.4 &   45.843 & M\\
   PtS$_2$ &  5.019 --  5.043 &  4.575 & -8.9 -- -9.3 & 49.0 &   21.599 &    0.112\\
   PtSe$_2$ &  5.031 --  5.082 &  4.797 & -4.6 -- -5.6 & 67.3 &   25.136 & M\\
   PtTe$_2$ &  5.201 --  5.224 &  5.086 & -2.2 -- -2.6 & 80.5 &   32.849 & M\\
   RhTe$_2$ &  5.410 --  5.442 &  5.197 & -3.9 -- -4.5 & 86.5 &   51.709 & M\\
   SnS$_2$ &  5.460 --  5.960 &  5.690 &  0.2 -- -4.5 & 30.0 &   11.235 &    1.113\\
   SnSe$_2$ &  6.128 --  6.141 &  5.910 & -3.6 -- -3.8 & 30.6 &   13.477 &    0.275\\
   SrFI &  8.888 --  8.916 &  7.859 & -11.6 -- -11.9 & 22.5 &   13.449 &    4.161\\
   TaS$_2$ &  5.853 --  5.900 &  5.691 & -2.8 -- -3.5 & 50.9 &   16.563 & M\\
   TaS$_2$ & 12.097 -- 12.100 & 11.804 & -2.4 -- -2.4 & 56.1 &   16.897 & M\\
   TaSe$_2$ &  6.203 --  6.272 &  5.998 & -3.3 -- -4.4 & 53.3 &   18.043 & M\\
   TaSe$_2$ & 12.696 -- 12.720 & 12.495 & -1.6 -- -1.8 & 60.7 &   18.396 & M\\
   TaSe$_2$ & 25.143 -- 25.500 & 25.147 & -0.0 -- -1.4 & 49.5 &   20.085 & M\\
   TiS$_2$ &  5.680 --  5.716 &  5.386 & -5.2 -- -5.8 & 63.9 &   18.651 & M\\
   TiSe$_2$ &  5.981 --  6.011 &  5.698 & -4.7 -- -5.2 & 69.0 &   20.279 & M\\
   TiTe$_2$ &  6.459 --  6.539 &  6.225 & -3.6 -- -4.8 & 67.1 &   24.132 & M\\
   TlCrTe$_2$ &  7.839 --  7.935 &  7.712 & -1.6 -- -2.8 & 77.6 &   70.366 & M\\
   VBr$_2$ &  6.206 &  5.715 & -7.9 & 29.1 &    9.775 & M\\
   VCl$_2$ &  5.798 --  5.835 &  5.330 & -8.1 -- -8.6 & 30.2 &    9.181 & M\\
   VI$_2$ &  6.714 &  6.148 & -8.4 & 27.9 &   10.462 & M\\
   VS$_2$ &  5.755 --  5.755 &  5.390 & -6.3 -- -6.3 & 55.4 &   21.171 & M\\
   VSe$_2$ &  6.048 --  6.150 &  5.770 & -4.6 -- -6.2 & 62.0 &   20.504 & M\\
   WS$_2$ & 12.323 -- 12.500 & 12.230 & -0.8 -- -2.2 & 51.3 &   12.898 &    0.924\\
   WS$_2$ & 18.490 & 18.288 & -1.1 & 45.1 &   12.549 &    0.759\\
   WSe$_2$ & 12.960 -- 12.980 & 12.871 & -0.7 -- -0.8 & 52.5 &   13.355 &    0.996\\
   Y$_2$I$_2$Ga$_2$ & 11.434 & 10.968 & -4.1 & 30.9 &    8.443 & M\\
   YI$_3$ & 20.880 & 20.528 & -1.7 & 20.0 &    7.868 &    2.450\\
   ZrNCl & 27.178 -- 27.672 & 27.148 & -0.1 -- -1.9 & 41.6 &    6.975 &    1.801\\
   ZrS$_2$ &  5.810 --  5.850 &  5.626 & -3.2 -- -3.8 & 42.7 &   12.681 &    0.536\\
   ZrSe$_2$ &  6.125 --  6.192 &  5.912 & -3.5 -- -4.5 & 46.7 &   14.678 & M\\
   ZrTe$_2$ &  6.630 --  6.660 &  6.403 & -3.4 -- -3.9 & 62.9 &   20.929 & M\\
\end{longtable}

\newpage
\subsection{PBE-D}
\begin{longtable}{l|cccccc}
 Name & $c_{exp}$ [\AA] & $c_{calc}$ & $c$ dev. [\%] & $C_{33}$ [GPa] & $E_B$ [meV/\AA$^2$] & $E_g$ [eV] \\\hline\hline
   AgBiP$_2$Se$_6$ & 39.615 & 40.029 &  1.0 & 23.0 &    8.630 &    1.246\\
   BBr$_3$ &  6.847 --  6.864 &  6.757 & -1.3 -- -1.6 & 13.9 &    8.768 &    3.563\\
   BI$_3$ &  7.261 --  7.460 &  7.015 & -3.4 -- -6.0 & 21.9 &   12.027 &    2.545\\
   BN &  6.661 --  6.690 &  6.477 & -2.8 -- -3.2 & 69.1 &   28.057 &    4.146\\
   BaFI &  7.962 --  8.102 &  8.059 & -0.5 --  1.2 & 103.4 &   44.343 &    3.869\\
   Bi$_2$Te$_3$ & 30.440 -- 30.497 & 30.785 &  0.9 --  1.1 & 48.3 &   25.628 &    0.783\\
   C &  6.704 --  6.930 &  6.740 &  0.5 -- -2.7 & 44.0 &   21.148 & M\\
   CdI$_2$ &  6.835 --  6.864 &  6.808 & -0.4 -- -0.8 & 33.2 &   14.652 &    2.268\\
   CoTe$_2$ &  5.405 &  5.447 &  0.8 & 40.8 &   53.724 & M\\
   CrSe$_2$ &  5.915 &  6.022 &  1.8 & 40.8 &   22.541 & M\\
   Cu$_2$S &  6.670 --  6.680 &  6.175 & -7.4 -- -7.6 & 54.6 &   52.033 & M\\
   FeLiAs &  6.349 &  5.782 & -8.9 & 209.2 &   80.748 & M\\
   FeS &  5.039 &  4.443 & -11.8 & 35.4 &   24.178 & M\\
   GaS & 14.230 -- 15.530 & 15.632 &  0.7 --  9.9 & 31.4 &   12.967 &    1.853\\
   GaSe & 15.919 -- 15.995 & 15.881 & -0.2 -- -0.7 & 51.5 &   15.768 &    1.465\\
   Ge$_2$Sb$_2$Te$_5$ & 16.960 -- 17.239 & 17.472 &  1.4 --  3.0 & 43.2 &   24.107 &    0.119\\
   HfS$_2$ &  5.837 --  5.856 &  6.043 &  3.2 --  3.5 & 21.6 &    7.657 &    1.223\\
   HfSe$_2$ &  6.143 --  6.159 &  6.029 & -1.9 -- -2.1 & 54.6 &   30.752 &    0.337\\
   HfTe$_2$ &  6.650 --  6.670 &  6.678 &  0.1 --  0.4 & 56.3 &   25.983 & M\\
   HgI$_2$ & 12.088 -- 12.450 & 12.366 & -0.6 --  2.3 & 25.7 &   16.524 &    1.108\\
   In$_2$Zn$_2$S$_5$ & 46.270 & 45.957 & -0.7 & 100.8 &   45.257 &    0.361\\
   Mg$_2$(P$_2$Se$_6$) & 20.194 & 20.173 & -0.1 & 34.9 &   12.561 &    2.025\\
   MgBr$_2$ &  6.260 --  6.269 &  6.343 &  1.2 --  1.3 & 24.6 &    9.801 &    4.563\\
   MgI$_2$ &  6.862 --  6.895 &  6.822 & -0.6 -- -1.1 & 30.2 &   12.222 &    3.482\\
   MoS$_2$ & 12.290 -- 12.530 & 12.389 &  0.5 -- -1.1 & 51.2 &   18.211 &    0.970\\
   MoS$_2$ & 18.330 -- 18.450 & 18.469 &  0.1 --  0.8 & 53.0 &   19.032 &    1.187\\
   MoSe$_2$ & 12.900 -- 12.930 & 13.012 &  0.6 --  0.9 & 69.1 &   24.158 &    1.024\\
   MoTe$_2$ & 13.964 -- 13.974 & 14.025 &  0.4 --  0.4 & 85.6 &   31.969 &    0.885\\
   NbS$_2$ & 17.800 -- 17.918 & 17.859 &  0.3 --  0.3 & 51.2 &   18.050 & M\\
   NbSe$_2$ & 12.482 -- 12.550 & 12.691 &  1.1 --  1.7 & 50.9 &   25.222 & M\\
   NbTe$_2$ &  6.610 &  6.877 &  4.0 & 61.3 &   31.169 & M\\
   Ni$_2$SbTe$_2$ & 15.634 -- 15.682 & 15.673 & -0.1 --  0.3 & 33.3 &   40.323 & M\\
   NiSbSi &  8.179 &  7.813 & -4.5 & 57.4 &   75.287 & M\\
   NiTe$_2$ &  5.251 --  5.308 &  5.216 & -0.7 -- -1.7 & 50.1 &   48.955 & M\\
   PdTe$_2$ &  5.113 --  5.270 &  5.114 &  0.0 -- -3.0 & 84.2 &   54.430 & M\\
   PtS$_2$ &  5.019 --  5.043 &  5.457 &  8.2 --  8.7 & 21.3 &   10.301 &    1.043\\
   PtSe$_2$ &  5.031 --  5.082 &  5.471 &  7.7 --  8.7 & 13.4 &   19.306 & M\\
   PtTe$_2$ &  5.201 --  5.224 &  5.190 & -0.2 -- -0.7 & 62.2 &   41.530 & M\\
   RhTe$_2$ &  5.410 --  5.442 &  5.348 & -1.2 -- -1.7 & 55.4 &   59.688 & M\\
   SnS$_2$ &  5.460 --  5.960 &  5.962 &  0.0 --  9.2 & 30.2 &   14.109 &    1.472\\
   SnSe$_2$ &  6.128 --  6.141 &  6.221 &  1.3 --  1.5 & 42.0 &   18.703 &    0.535\\
   SrFI &  8.888 --  8.916 &  8.473 & -4.7 -- -5.0 & 22.0 &   17.121 &    4.337\\
   TaS$_2$ &  5.853 --  5.900 &  6.164 &  4.5 --  5.3 & 33.8 &   11.188 & M\\
   TaS$_2$ & 12.097 -- 12.100 & 12.209 &  0.9 --  0.9 & 33.8 &   11.080 & M\\
   TaSe$_2$ &  6.203 --  6.272 &  6.385 &  1.8 --  2.9 & 55.3 &   19.641 & M\\
   TaSe$_2$ & 12.696 -- 12.720 & 12.889 &  1.3 --  1.5 & 57.7 &   20.057 & M\\
   TaSe$_2$ & 25.143 -- 25.500 & 25.483 & -0.1 --  1.4 & 86.4 &   22.251 & M\\
   Ti$_2$PTe$_2$ & 28.486 & 28.375 & -0.4 & 96.1 &   27.842 & M\\
   TiS$_2$ &  5.680 --  5.716 &  5.832 &  2.0 --  2.7 & 40.7 &   14.371 & M\\
   TiSe$_2$ &  5.981 --  6.011 &  6.120 &  1.8 --  2.3 & 42.8 &   19.553 & M\\
   TiTe$_2$ &  6.459 --  6.539 &  6.609 &  1.1 --  2.3 & 59.7 &   28.211 & M\\
   VBr$_2$ &  6.206 &  6.095 & -1.8 & 34.0 &   13.654 & M\\
   VCl$_2$ &  5.798 --  5.835 &  5.777 & -0.4 -- -1.0 & 22.8 &   14.091 & M\\
   VI$_2$ &  6.714 &  6.488 & -3.4 & 41.1 &   17.363 & M\\
   VS$_2$ &  5.755 --  5.755 &  5.905 &  2.6 --  2.6 & 44.5 &   17.071 & M\\
   VSe$_2$ &  6.048 --  6.150 &  6.266 &  1.9 --  3.6 & 55.1 &   22.468 & M\\
   WS$_2$ & 12.323 -- 12.500 & 12.523 &  0.2 --  1.6 & 34.0 &   11.042 &    1.307\\
   WS$_2$ & 18.490 & 18.656 &  0.9 & 30.6 &   10.698 &    1.333\\
   WSe$_2$ & 12.960 -- 12.980 & 13.176 &  1.5 --  1.7 & 65.2 &   19.936 &    1.173\\
   Y$_2$I$_2$Ga$_2$ & 11.434 & 11.358 & -0.7 & 47.9 &   15.947 & M\\
   YS$_2$ &  7.846 &  7.952 &  1.4 & 172.4 &   80.072 &    0.404\\
   ZrNCl & 27.178 -- 27.672 & 27.629 & -0.2 --  1.7 & 54.8 &   12.785 &    1.818\\
   ZrS$_2$ &  5.810 --  5.850 &  5.907 &  1.0 --  1.7 & 33.5 &   12.739 &    0.817\\
   ZrSe$_2$ &  6.125 --  6.192 &  6.215 &  0.4 --  1.5 & 47.1 &   17.909 &    0.246\\
   ZrTe$_2$ &  6.630 --  6.660 &  6.714 &  0.8 --  1.3 & 60.1 &   26.737 & M\\
\end{longtable}

\newpage
\subsection{PBE}
\begin{longtable}{l|cccccc}
 Name & $c_{exp}$ [\AA] & $c_{calc}$ & $c$ dev. [\%] & $C_{33}$ [GPa] & $E_B$ [meV/\AA$^2$] & $E_g$ [eV] \\\hline\hline
   AgBiP$_2$Se$_6$ & 39.615 & 40.427 &  2.0 & 1.9 &    0.568 &    1.377\\
   BBr$_3$ &  6.847 --  6.864 &  7.728 & 12.6 -- 12.9 & 0.9 &    0.616 &    3.641\\
   BI$_3$ &  7.261 --  7.460 &  8.062 &  8.1 -- 11.0 & 1.0 &    0.556 &    2.542\\
   BN &  6.661 --  6.690 &  7.686 & 14.9 -- 15.4 & 1.1 &    0.571 &    4.409\\
   BaFI &  7.962 --  8.102 &  8.331 &  2.8 --  4.6 & 16.0 &    4.326 &    4.078\\
   BiIO &  9.128 --  9.151 & 10.509 & 14.8 -- 15.1 & 2.1 &    0.569 &    2.067\\
   C &  6.704 --  6.930 &  7.809 & 12.7 -- 16.5 & 1.2 &    0.399 & M\\
   CdI$_2$ &  6.835 --  6.864 &  7.937 & 15.6 -- 16.1 & 1.5 &    0.448 &    2.366\\
   CoTe$_2$ &  5.405 &  5.560 &  2.9 & 42.2 &   21.750 & M\\
   CrSe$_2$ &  5.915 &  6.699 & 13.3 & 3.2 &    0.619 & M\\
   CrSiTe$_3$ & 20.528 -- 20.582 & 21.448 &  4.2 --  4.5 & 2.4 &    0.914 &    0.427\\
   Cu$_2$S &  6.670 --  6.680 &  7.091 &  6.2 --  6.3 & 4.6 &    2.165 & M\\
   Fe(PSe$_3$) & 19.800 -- 19.812 & 20.890 &  5.4 --  5.5 & 2.0 &    0.544 & M\\
   GaS & 14.230 -- 15.530 & 16.721 &  7.7 -- 17.5 & 2.1 &    0.747 &    2.377\\
   GaSe & 15.919 -- 15.995 & 17.261 &  7.9 --  8.4 & 1.8 &    0.442 &    2.161\\
   Ge$_2$Sb$_2$Te$_5$ & 16.960 -- 17.239 & 18.665 &  8.3 -- 10.1 & 3.6 &    1.475 &    0.177\\
   HfS$_2$ &  5.837 --  5.856 &  6.855 & 17.1 -- 17.4 & 2.1 &    0.760 &    1.317\\
   HfSe$_2$ &  6.143 --  6.159 &  7.086 & 15.0 -- 15.3 & 1.9 &    0.706 &    0.599\\
   HfTe$_2$ &  6.650 --  6.670 &  7.210 &  8.1 --  8.4 & 8.0 &    1.989 & M\\
   HgI$_2$ & 12.088 -- 12.450 & 13.550 &  8.8 -- 12.1 & 2.0 &    0.784 &    1.391\\
   Mg$_2$(P$_2$Se$_6$) & 20.194 & 21.178 &  4.9 & 1.6 &    0.481 &    2.035\\
   MgBr$_2$ &  6.260 --  6.269 &  7.329 & 16.9 -- 17.1 & 1.7 &    0.598 &    4.655\\
   MgI$_2$ &  6.862 --  6.895 &  7.927 & 15.0 -- 15.5 & 1.7 &    0.415 &    3.617\\
   MoS$_2$ & 12.290 -- 12.530 & 13.521 &  7.9 -- 10.0 & 1.5 &    0.714 &    1.673\\
   MoS$_2$ & 18.330 -- 18.450 & 19.723 &  6.9 --  7.6 & 1.5 &    0.598 &    1.745\\
   MoSe$_2$ & 12.900 -- 12.930 & 14.099 &  9.0 --  9.3 & 1.1 &    0.397 &    1.468\\
   MoTe$_2$ & 13.964 -- 13.974 & 14.671 &  5.0 --  5.1 & 4.0 &    1.248 &    1.044\\
   NbS$_2$ & 17.800 -- 17.918 & 18.418 &  2.8 --  3.5 & 6.8 &    1.664 & M\\
   NbSe$_2$ & 12.482 -- 12.550 & 13.172 &  5.0 --  5.5 & 6.9 &    1.522 & M\\
   NbSe$_2$ & 25.230 -- 25.450 & 25.511 &  0.2 --  1.1 & 10.5 &    2.643 & M\\
   NbTe$_2$ &  6.610 &  7.330 & 10.9 & 7.5 &    2.349 & M\\
   Ni$_2$SbTe$_2$ & 15.634 -- 15.682 & 15.971 &  1.8 --  2.2 & 27.7 &    5.857 & M\\
   NiSbSi &  8.179 &  7.760 & -5.1 & 48.5 &   26.875 & M\\
   NiTe$_2$ &  5.251 --  5.308 &  5.393 &  1.6 --  2.7 & 45.7 &   17.068 & M\\
   PbBi$_4$Te$_7$ & 23.600 -- 23.892 & 24.785 &  3.7 --  5.0 & 2.8 &    1.088 &    0.937\\
   PbFI &  8.770 --  8.800 & 10.347 & 17.6 -- 18.0 & 1.7 &    0.461 &    2.324\\
   PbO &  4.988 --  5.071 &  5.779 & 14.0 -- 15.9 & 6.9 &    2.983 &    1.945\\
   PbSb$_2$Te$_4$ & 41.712 & 42.434 &  1.7 & 1.3 &    0.724 &    0.580\\
   PdTe$_2$ &  5.113 --  5.270 &  5.298 &  0.5 --  3.6 & 69.9 &   18.314 & M\\
   PtS$_2$ &  5.019 --  5.043 &  6.581 & 30.5 -- 31.1 & 1.6 &    0.741 &    1.509\\
   PtSe$_2$ &  5.031 --  5.082 &  6.651 & 30.9 -- 32.2 & 1.5 &    0.626 &    0.921\\
   PtTe$_2$ &  5.201 --  5.224 &  5.473 &  4.8 --  5.2 & 33.1 &    5.416 & M\\
   Re(AgCl$_3$)$_2$ & 16.731 -- 16.731 & 17.109 &  2.3 --  2.3 & 8.9 &    1.297 & M\\
   RhTe$_2$ &  5.410 --  5.442 &  5.544 &  1.9 --  2.5 & 48.6 &   25.649 & M\\
   SnS$_2$ &  5.460 --  5.960 &  7.068 & 18.6 -- 29.4 & 1.9 &    0.808 &    1.551\\
   SnSe$_2$ &  6.128 --  6.141 &  7.209 & 17.4 -- 17.6 & 2.1 &    0.734 &    0.668\\
   SrFI &  8.888 --  8.916 &  9.876 & 10.8 -- 11.1 & 2.5 &    0.708 &    4.323\\
   TaS$_2$ &  5.853 --  5.900 &  6.894 & 16.8 -- 17.8 & 3.0 &    1.126 & M\\
   TaS$_2$ & 12.097 -- 12.100 & 13.139 &  8.6 --  8.6 & 2.3 &    1.029 & M\\
   TaSe$_2$ &  6.203 --  6.272 &  6.904 & 10.1 -- 11.3 & 6.0 &    1.519 & M\\
   TaSe$_2$ & 12.696 -- 12.720 & 13.534 &  6.4 --  6.6 & 3.5 &    1.065 & M\\
   TaSe$_2$ & 25.143 -- 25.500 & 26.317 &  3.2 --  4.7 & 6.9 &    2.133 & M\\
   Ti$_2$PTe$_2$ & 28.486 & 28.702 &  0.8 & 13.9 &    4.008 & M\\
   TiS$_2$ &  5.680 --  5.716 &  6.589 & 15.3 -- 16.0 & 2.3 &    1.075 & M\\
   TiSe$_2$ &  5.981 --  6.011 &  6.728 & 11.9 -- 12.5 & 4.6 &    1.067 & M\\
   TiTe$_2$ &  6.459 --  6.539 &  6.856 &  4.8 --  6.1 & 9.7 &    2.799 & M\\
   TlCrTe$_2$ &  7.839 --  7.935 &  8.036 &  1.3 --  2.5 & 60.6 &   48.774 & M\\
   VBr$_2$ &  6.206 &  7.077 & 14.0 & 2.0 &    0.815 & M\\
   VCl$_2$ &  5.798 --  5.835 &  6.674 & 14.4 -- 15.1 & 2.3 &    0.753 & M\\
   VI$_2$ &  6.714 &  7.482 & 11.4 & 2.2 &    0.644 & M\\
   VS$_2$ &  5.755 --  5.755 &  6.636 & 15.3 -- 15.3 & 3.4 &    1.310 & M\\
   VSe$_2$ &  6.048 --  6.150 &  6.899 & 12.2 -- 14.1 & 4.7 &    1.237 & M\\
   WS$_2$ & 12.323 -- 12.500 & 13.657 &  9.3 -- 10.8 & 1.3 &    0.626 &    1.845\\
   WS$_2$ & 18.490 & 19.956 &  7.9 & 1.4 &    0.608 &    1.838\\
   WSe$_2$ & 12.960 -- 12.980 & 14.203 &  9.4 --  9.6 & 1.4 &    0.431 &    1.550\\
   Y$_2$I$_2$Ga$_2$ & 11.434 & 12.444 &  8.8 & 3.5 &    0.531 & M\\
   YI$_3$ & 20.880 & 21.476 &  2.9 & 2.0 &    0.524 &    2.804\\
   ZrNCl & 27.178 -- 27.672 & 28.534 &  3.1 --  5.0 & 3.5 &    0.539 &    1.931\\
   ZrS$_2$ &  5.810 --  5.850 &  6.779 & 15.9 -- 16.7 & 2.1 &    0.877 &    1.061\\
   ZrSe$_2$ &  6.125 --  6.192 &  6.917 & 11.7 -- 12.9 & 2.7 &    0.883 &    0.362\\
   ZrTe$_2$ &  6.630 --  6.660 &  7.005 &  5.2 --  5.7 & 8.4 &    2.950 & M\\
\end{longtable}

%